\documentclass[journal]{IEEEtran}
\usepackage{array}
\usepackage{marvosym}
\usepackage[cmex10]{amsmath}
\usepackage{amssymb,amsthm,epsfig}
\usepackage{graphpap}
\usepackage[latin1]{inputenc}
\usepackage{rotating}
\usepackage{algorithm}
\usepackage{setspace}
\usepackage{subfigure}
\usepackage{environ}
\usepackage{multirow}
\usepackage{setspace}
\usepackage{textcomp}
\usepackage[flushleft]{threeparttable}
\usepackage{color}
\usepackage{soul}

\usepackage{epsfig}
\usepackage{amsmath}
\usepackage{algcompatible}
\usepackage{latexsym}
\usepackage{amsfonts}
\usepackage[latin1]{inputenc}
\usepackage{rotating}
\usepackage{multirow}
\usepackage{setspace}
\usepackage{textcomp}
\usepackage[pdftex]{hyperref}
\usepackage{bm}
\hypersetup{colorlinks=true, allcolors=black, linkcolor=black, citecolor=black }
\usepackage{mathtools}

\usepackage{mdwlist}
\usepackage{graphicx}
\usepackage{cite}
\usepackage{enumitem}

\usepackage{footnote}
\makesavenoteenv{tabular}
\usepackage{tabulary}
\usepackage[table]{xcolor}
\newcolumntype{P}[1]{>{\centering\arraybackslash}p{#1}}
\newcolumntype{M}[1]{>{\centering\arraybackslash}m{#1}}
\begin{document}
	\title{Switching Device-Cognizant Sequential Distribution System Restoration}
	%
	\author{Anmar Arif,~\IEEEmembership{Member,~IEEE,}
		Bai Cui,~\IEEEmembership{Member,~IEEE,}
		Zhaoyu Wang,~\IEEEmembership{Senior Member,~IEEE,}
		\\
		\thanks{Z. Wang's work was supported the U.S. Department of Energy Wind Energy Technologies Office under Grant DE-EE0008956 (Corresponding author: Zhaoyu Wang).}
		\thanks{A. Arif is with the Department of Electrical Engineering, King Saud University, Riyadh, 11451 Saudi Arabia (E-mail: anarif\MVAt ksu.edu.sa)}
		\thanks{Z. Wang is with the Department of Electrical and Computer Engineering, Iowa State University, Ames, IA, 50011. (Email: wzy\MVAt iastate.edu)}
		\thanks{B. Cui is with the National Renewable Energy Laboratory, Golden, CO 80401, USA. (Email: bcui\MVAt nrel.gov)}
	}
	\maketitle
	\begin{abstract}
		This paper presents an optimization framework for sequential reconfiguration using an assortment of switching devices and repair process in distribution system restoration. Compared to existing studies, this paper considers types, capabilities and operational limits of different switching devices, making it applicable in practice. We develop a novel multi-phase method to find the optimal sequential operation of various switching devices and repair faulted areas. We consider circuit breakers, reclosers, sectionalizers, load breaker switches, and fuses. The switching operation problem is decomposed into two mixed-integer linear programming (MILP) subproblems. The first subproblem determines the optimal network topology and estimates the number of steps to reach that topology, while the second subproblem generates a sequence of switching operations to coordinate the switches. For repairing the faults, we design an MILP model that dispatches repair crews to clear faults and replace melted fuses. After clearing a fault, we update the topology of the network by generating a new sequence of switching operations, and the process continues until all faults are cleared. To improve the computational efficiency, a network reduction algorithm is developed to group line sections, such that only switchable sections are present in the reduced network. 
		The proposed method is validated on the IEEE 123-bus and 8500-bus systems.
	\end{abstract}
	
	\begin{IEEEkeywords}
		Distribution system, integer programming, fault isolation, service restoration
	\end{IEEEkeywords}
	\vspace{-0.2cm}
	\section*{Nomenclature}
	{
		\small
		\addcontentsline{toc}{section}{Nomenclature}
		\begin{description}[style=multiline,leftmargin=3cm]
			\item[\textbf{Sets and Indices}]
		\end{description}
		\begin{description}[style=multiline,leftmargin=1.5cm,itemsep=0.05cm] 
			\item[$i/j$] Indices for buses and bus blocks
			\item[$k/l$] Index for distribution line connecting $i$ and $j$
			\item[$s$] Index for step number
			\item[$\varphi$] Index for phase number
			\item[$\Omega_{B},\Omega_{BL}$] Set of buses and set of bus blocks
			\item[$\Omega_{CB}$] Set of circuit breakers and reclosers
			\item[$\Omega_{DB}$] Set of bus blocks that contain damaged components
			\item[$\Omega_{F},\Omega_{F(i)}$] Set of faulted lines and set of faulted lines in bus block $i$
			\item[$\Omega_{FS}$] Set of lines with fuses
			\item[$\Omega_{MF}$] Set of fuses that need replacement
			\item[$\Omega_{MS}$] Set of manual sectionalizing switches
			\item[$\Omega_{SW}$] Set of all switches including fuses
			\item[$\Omega_{Sub}$] Set of buses connected to substations or generators
			\item[$\Omega_{K}$] Set of lines
			\item[$\Omega_{K(.,i)}$] Set of lines with bus $i$ as the to bus
			\item[$\Omega_{K(i,.)}$] Set of lines with bus $i$ as the from bus
			\item[$\Omega_{LBS}$] Set of load breaker switches
			\item[$\Omega_{Sec}$] Set of sectionalizing switches
		\end{description}
		\begin{description}[style=multiline,leftmargin=3cm]
			\item[\textbf{Parameters}]
		\end{description}
		\begin{description}[style=multiline,leftmargin=1.5cm,itemsep=0.06cm]
			\item[$ET_{k}$] Repair time of line $k$
			\item[$\check{I}_k/\hat{I}_k$] Making/breaking current capacity of switch $k$
			\item[${p}_{k\varphi}$] Binary parameter indicating the presence of phase $\varphi$ at line $k$ 
			\item[$P^D_{i \varphi}/Q^D_{i \varphi}$] Active/reactive demand at bus $i$ and phase $\varphi$
			\item[$\tilde{P}^D_{i \varphi}/\tilde{Q}^D_{i \varphi}$] Aggregated active/reactive demand at bus block $i$ and phase $\varphi$
			\item[$\bar{S}_{k}$] Maximum apparent power for line $k$
			\item[$\bar{P}^G_i/\bar{Q}^G_i$] Maximum active/reactive power for generator $i$
			\item[$\mathcal{T}^S_k$] Operation time of switch $k$
			\item[${tr}_{k l}$] Travel time between manual switches $k$ and $l$
			\item[$\grave{tr}_{ij}$] Travel time between bus blocks $i$ and $j$
			\item[$\bar{w},\bar{\gamma}$] Maximum waiting time and number of switching actions
			\item[$\Gamma^0_{k}/\Gamma^F_{k}$] Binary parameter representing the initial/final state of switch $k$
			\item[$Z_k$] The impedance matrix of line $k$ 
			\item[$\rho^D_i,\rho^{SW}_k$] The cost of shedding per unit load at bus $i$ and the cost of switching
			\item[$\rho^T_{ij}$] Cost of traveling from location $i$ to $j$
			\item[$\rho^R$] Penalty cost for total switching operation time
		\end{description}
		\begin{description}[style=multiline,leftmargin=3cm]
			\item[\textbf{Decision Variables}]
		\end{description}
		\begin{description}[style=multiline,leftmargin=1.5cm,itemsep=0.08cm]
			\item[$\alpha_{kc}$] Arrival time at manual switch $k$ for crew $c$
			\item[$\grave{\alpha}_{ic}$] Arrival time at bus block $i$ for crew $c$
			\item[$\mathcal{O}_s$] The time elapsed after switching step $s$
			\item[$P_{k \varphi}/Q_{k \varphi}$] Active/reactive power flowing on line $k$ and phase $\varphi$
			\item[$P^G_{i \varphi}/Q^G_{i \varphi}$] Active/reactive power generated at bus $i$ and phase $\varphi$
			\item[$\gamma_{k s}$] Binary variable indicates whether switch $k$ is operated in step $s$
			\item[$\mathcal{R}_i$] The time when all damaged components in bus block $i$ are repaired
			\item[$w_k$] Crew wait time at manual switch $k$
			\item[$x_{k l c}$] Binary variable equal to 1 if crew $c$ travels from switch $k$ to $l$
			\item[$\grave{x}_{ijc}$] Binary variable equal to 1 if crew $c$ travels from bus block $i$ to $j$
			\item[$x^F_{i s}$] Binary variable equal to 1 if bus $i$ is in a faulted area in step $s$
			\item[$x^E_{i s}$] Binary variable equal to 1 if bus $i$ can be served by a generator
			\item[$u_{k s}$] Binary variable indicating the status of line $k$
			\item[$\mathcal{W}_{k c}$] Binary variable equal to 1 if crew c is assigned to damaged component $k$
			\item[$y_{i s}$] Connection status of the loads at bus $i$ and step $s$
			\item[${S}_{k\varphi,s}$] Apparent power of each phase for line $k$ at step $s$
			\item[${U}_{i\varphi}$] The squared voltage magnitude at bus $i$ for phase $\varphi$
			\item[$\mathcal{X}_{i t}$] Binary variable equal to 0 if bus $i$ is in an outage area at time $t$
		\end{description}
	}
	%
	%
	\section{Introduction}\label{sec:1}

	\IEEEPARstart{D}{istribution} {\color{black}networks are experiencing major changes with the development of smart grid technologies. Advanced control and measurement devices are being introduced to the network in order to have a resilient and more controllable system. The integration of automatic and remotely controllable switches with communication technologies allows the distribution system operator to quickly recover from anomalies and reduce the outage duration for the customers. 
		
		\vspace{-0.2cm}
		\subsection{Motivation}
		
		Once a distribution system is damaged, the faults in the system are isolated automatically using protective devices (e.g., reclosers and circuit breakers), and repair crews are then sent to clear the permanent faults. Meanwhile, some customers will likely lose power while the crews are repairing the faults. During this process, the distribution system operator will reconfigure the topology of the system through a sequence of switching operations, in order to restore service to as many customers as possible while keeping the faults isolated. Once a damaged section is repaired, the switches are operated again in order to restore the area. The switching operation in distribution systems involves the coordination of different switching devices such as circuit breakers (CB), reclosers (REC), sectionalizers (SEC), and load breaker switches (LBS). Due to the diverse kind of switching devices in the network and their different characteristics and limitations, the switches must be coordinated and operated in a specific sequence. CBs and RECs can be operated at any time. SECs can be operated at no-load only. LBSs can be operated under load (with specified current rating), but cannot make or interrupt fault currents. In addition, some switches can be controlled remotely, while others must be operated manually by field crews. Manually operated switches must be de-energized before crews can operate them to ensure their safety. Therefore, it is critical to develop an effective and fast method to find the sequence of switching operations.
		
		\vspace{-0.2cm}
		\subsection{Literature Review}
		
		There has been considerable progress in power system restoration techniques in distribution systems\cite{Y_Wang2016}. A variety of methods on distribution system restoration have been proposed, including microgrid formation \cite{Ding2017}, network reconfiguration using dynamic programming \cite{ChongWang2020}, and utilizing mobile resources \cite{Lei2019}. 
		Network reconfiguration is one of the most commonly used methods to restore a power distribution system. The authors in \cite{Butler2001} developed a reconfiguration formulation using a variation of the fixed charge network problem for service restoration. In \cite{Schmitz2021}, the authors developed an algorithm and a price-based mixed-integer linear program (MILP) model for co-optimizing the repair and operation of the distribution system, while considering energy storage and flexible loads. In \cite{Chen2016}, a MILP was formulated to maximize the critical loads to be served by operating remotely controlled switches to form microgrids. However, these methods consider network reconfiguration as a single step problem, where only the final topology is obtained. Multi-time step sequential methods are presented in \cite{Carvalho2007,Liu1998,Lopez2018,Li2014,B_Chen_multistep,B_Chen2018,Zhang2020}. In \cite{Liu1998}, the authors developed a rule-based expert system for finding the switching actions required to restore customers affected by an outage. Restoration was accomplished by heuristically finding a plan to restore as many customers as possible following a set of predefined rules. The authors in \cite{Carvalho2007} used a two-step approach for post-fault restoration. The first-step used Genetic Algorithm to find the optimal topology, and the second step used Dynamic Programming to find the sequence of operations. In \cite{Li2014}, the authors developed a graph-theoretic method for restoring unbalanced distribution systems with distributed generators. The authors used the spanning tree search algorithm to find the sequence of switching operations, where the objective was to minimize the number of switching steps and maximize the restored load. The paper in \cite{Lopez2018} developed mixed-integer nonlinear programming (MINLP) and MILP models for solving the restoration problem and obtain the switching sequence. The authors included constraints on the maximum current through a switch, but did not consider the breaking and making capacities of the switches. Reference \cite{B_Chen_multistep} developed a multi-time-step MILP formulation for service restoration. The authors continued their work in \cite{B_Chen2018}, where the sequential operation was applied considering unbalanced power operations. However, \cite{B_Chen_multistep} and \cite{B_Chen2018} assumed all switches and loads are disconnected in the initial step. In \cite{Zhang2020}, the authors presented a study for optimizing the operation of manual and remotely controlled switches, in addition to optimizing the repair process of the damaged components in balanced distribution networks.   
		
		\subsection{Contribution}
		
		The previous studies assumed switching devices were uniform in distribution grids and neglected their different operational capabilities, which does not reflect the behaviour of the switches in distribution system restoration and could lead to infeasible switching operations. Sequential service restoration with the coordination of different types of switches is a challenging problem. The difficulties lie partly in modeling the intricate coordination between switches and their interactions with other components in the distribution system. Moreover, the required number of switching operations to reach the final optimal topology is unknown beforehand; addressing this challenge by brute-force trials or dynamic programming is infeasible since the problem must be solved in a short time. To the best of our knowledge, the proposed methodology is the first to consider the characteristics of switches and derive feasible sequence of operations in a systematic and mathematically rigorous manner. The contributions of this paper are listed below:
		
		\begin{itemize}
			\item We develop an optimization framework that assists decision makers to repair and restore distribution systems after permanent faults.
			\item We develop a new MILP model to solve the sequential switching problem in distribution system restoration.
			\item We model the characteristics and behaviour of different types of switches and their interactions in the sequential switching operation.
			\item We exploited the special problem structure and developed preprocessing techniques and problem simplifications tailored for the sequential restoration problem, such as using the concept of bus block, and estimating the maximum number of switching operations.
		\end{itemize}

		
		
	}
	
	The rest of this paper is organized as follows. Section II presents the proposed methodology and problem formulation. Section III presents the simulation results and Section IV concludes this paper.

	\section{Switching Device-Cognizant Restoration}\label{sec:2}
	
\begin{figure}[h!]
		\setlength{\abovecaptionskip}{0pt} 
		\setlength{\belowcaptionskip}{0pt} 
		\centering
		\includegraphics[width=0.46\textwidth]{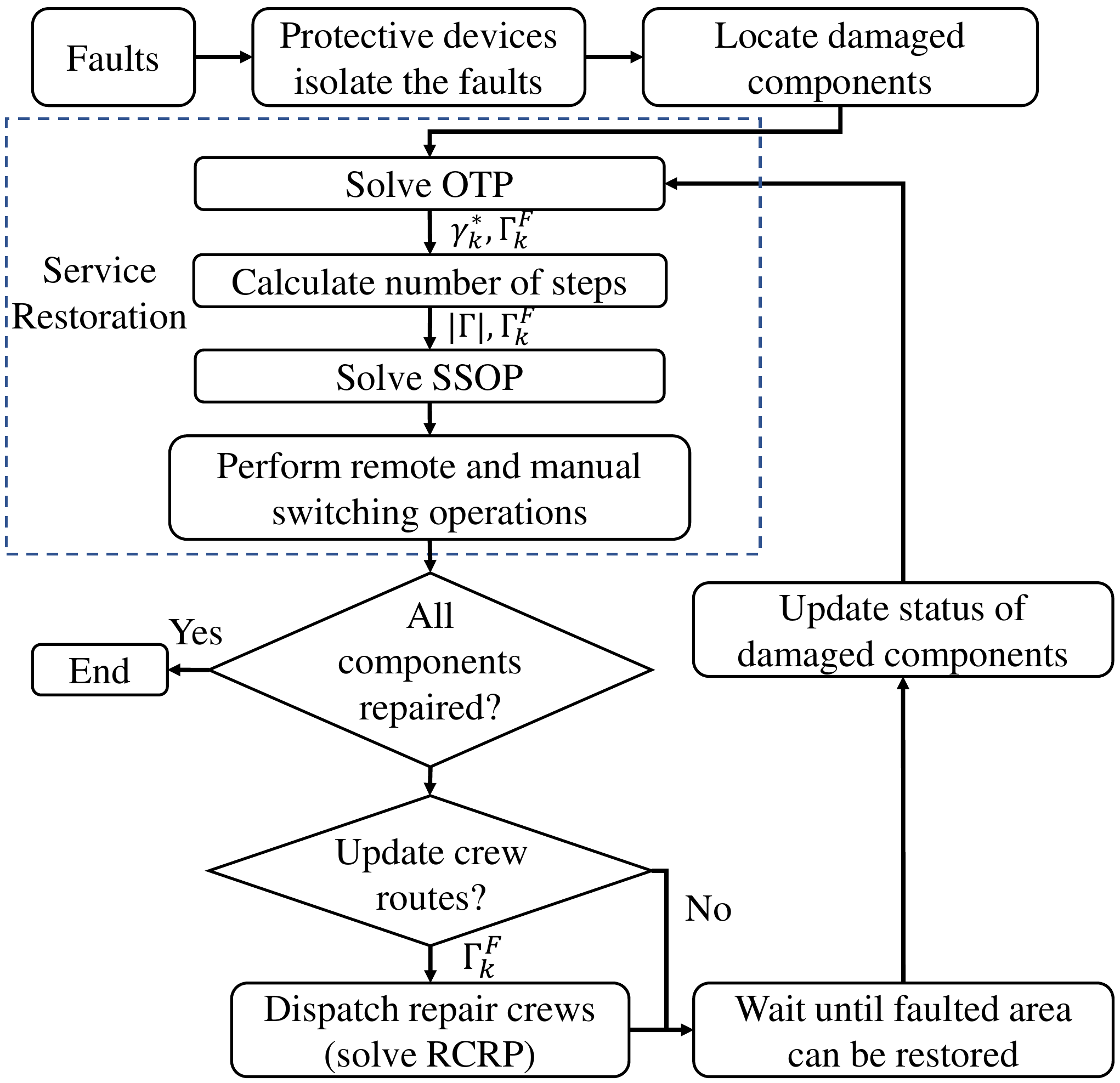}
		\caption{Flowchart of the service restoration approach.}\label{Flowchart}
	\end{figure}

	In this paper, we develop a multi-time step methodology to find the optimal sequential switching operation. Fig. \ref{Flowchart} depicts the methodology we employ for repair and service restoration. 
	
	When a distribution system experiences faults, protective devices will operate automatically to isolate the faults {\color{black}(readers can refer to \cite{Baghaee2018} for a study on distribution system protection and relay coordination)}. Damage assessors are then dispatched to locate the exact location of the damaged components and assess the damage. We then perform service restoration by solving two MILP subproblems, the optimal topology problem (OTP) and the sequential switching operation problem (SSOP). OTP determines the final optimal network topology using a single time step model, and outputs the operation status $\gamma_k^*$ ($\gamma_k^* = 1$ if switch $k$ is operated) and the on/off status $\Gamma_k^F$ for each switch. We use the results obtained from OTP to estimate an upper bound for the number of switching operations ($|\Gamma|$). Selecting the number of switching steps before solving SSOP is critical in order to avoid infeasibility and long computation times \cite{Carvalho2007,B_Chen2018}. After setting the number of steps to $|\Gamma|$, we solve SSOP to generate the optimal sequence of switching operations for remotely and manually operated switches. The next step is the repair crew routing problem (RCRP). RCRP obtains the status of each switch ($\Gamma^F_k$) from OTP and SSOP, and then dispatches crews to clear faults and replace melted fuses. 
	Once crews repair a section of the network, the operator updates the operation and topology of the network by solving OTP and SSOP again. The process continues until all lines are repaired and all loads are restored. 

	\subsection{Switching Devices Modeling and Coordination}
	
	{\color{black} The switching devices in the distribution network can be categorized into three groups when it comes to restoration, the properties of which are summarized in Table \ref{table_switches}. In addition, each switch will have current breaking and making capacities. 
	We use CB, REC, LBS, and SEC, as examples of the different types of switches. RECs differ from CBs in that they are capable of automatically resetting if the excessive current ceases, in addition to being less expensive, lighter, and have lower short circuit ratings. In this paper, RECs are treated similarly to CBs since we tackle the restoration problem which is after the automatic operation of switches (fault isolation).}
	
	\begin{table}[htbp]
		\centering
		{\color{black}
			\caption{Types of switching devices for restoration}
			\vspace{-0.2cm}
			\begin{tabular}{|c|p{6.2cm}|p{1.2cm}|}
				\hline
				Type & Capabilities & Switches\\
				\hline
				1 & A switching device capable of making, carrying and breaking currents under normal and abnormal circuit conditions. & CB, REC \\
				\hline
				2 & Switches that can make or break current under normal load conditions, but cannot make or break fault currents. & LBS\\
				\hline
				3 & Switches that can be operated only under no-load conditions. & SEC \\
				\hline
			\end{tabular}%
			\label{table_switches}%
		}
	\end{table}%

\begin{figure}[h!]
		\setlength{\abovecaptionskip}{0pt} 
		\setlength{\belowcaptionskip}{0pt} 
		\centering
		\includegraphics[width=0.34\textwidth]{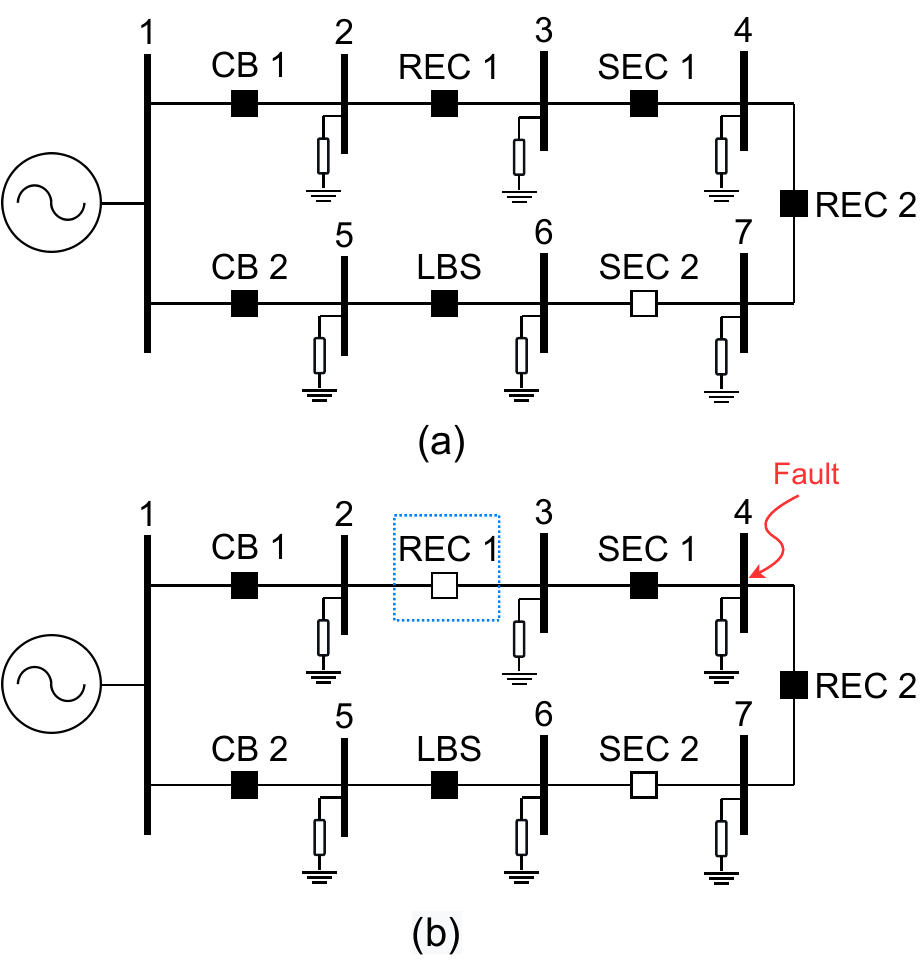}
		\caption{7-bus distribution system, where (a) is the default state and (b) is the initial state of switches after a fault near bus 4.}\label{Section_Example}
	\end{figure}

	An example is given that demonstrates the switching operations involved in the service restoration process. {\color{black} Consider the distribution system shown in Fig. \ref{Section_Example}, where (a) is the default state of the network and (b) is the initial state of switches after a fault near bus 4 occurs and REC 1 is operated automatically to isolate the fault.} 
	
\begin{figure}[h!]
		\setlength{\abovecaptionskip}{0pt} 
		\setlength{\belowcaptionskip}{0pt} 
		\centering
		\includegraphics[width=0.49\textwidth]{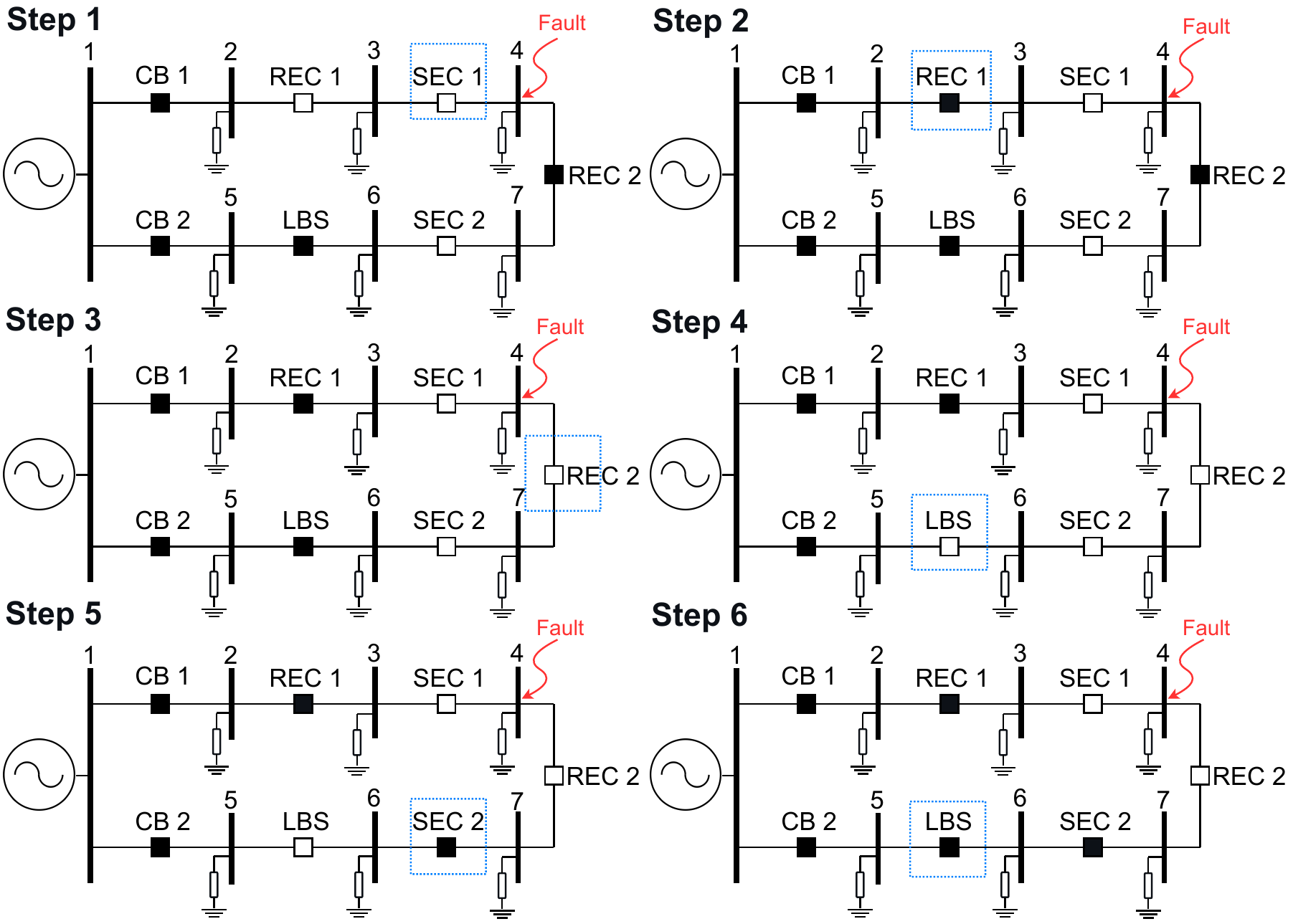}
		\caption{Optimal sequential switching for distribution system restoration.}\label{Section_Example_sol}
	\end{figure}

	\noindent
	The aim of the operator is to minimize the area that is affected by the fault through a sequence of switching operations. Therefore, SEC 1 and REC 2 should be opened, and all other switches closed to serve as many loads as possible. The steps taken to achieve the optimal topology are shown in Fig. \ref{Section_Example_sol}. SEC 1 is opened in the first step and REC 1 is closed in the second step. Once REC 1 is closed, the load at bus 3 can be served. In Step 3, REC 2 is opened to isolate bus 4. Next, SEC 2 must be closed to serve the load at bus 7, however, SEC 2 cannot be closed since bus 6 is energized. Therefore, the LBS is first opened and SEC 2 can then be closed. Finally, the LBS can then be closed in the final step. Subsequently, all loads can be served except the load at bus 4. It is seen that the entire process involves six steps even though only two switches change their statuses in the final topology. Multiple operation of the same switch may occurs due to limitations of some of the switches. However, a sectionalizer will not operate more than once in a switching sequence due to its limited operation capability. 
	In this paper, we assume that all CBs, RECs, and LBSs are remotely controllable, while some of the SECs are manual.
	
	\subsection{Calculating Final Optimal Topology}
	
	Before modeling the sequential switching problem, we first estimate the required number of switching steps. The study in \cite{B_Chen2018} selected the number of steps randomly and showed that by increasing it, the computation time rises exponentially. On the other hand, selecting a low number of steps could lead to an infeasible problem. In this paper, we first determine the final optimal topology by solving a single time step model, and then derive an equation for selecting the number of steps. 
	The mathematical model for OTP is given as follows: 
	
	{
		\[
		\text{\textbf{min} \{load shedding costs + switching costs\} 
		}
		\]
		\[
		\text{\textbf{subject to}} 
		\begin{cases}
		\text{Unbalanced power flow}
		\\
		\text{Switching and fault isolation}
		\\
		\text{Radiality~constraints}
		\end{cases}
		\]
	}\noindent	
	
	\noindent
	The detailed formulation can be found in Appendix A. The status of lines and switches are represented by a binary variable $u_k$. If a switch changes its status from open to close or vice versa, we use the binary variable $\gamma_{k}$ to represent this change of status. After solving OTP, we obtain the status of each switch $u_k^*$ and their operation status $\gamma_{k}^*$. The status of each switch is stored in $\Gamma^F_k=u^*_k$. Next, we calculate an upper bound ($|\Gamma|$) on the number of steps using $\gamma_{k}^*$. For each step, only one switching operation is made. 
	The variable $\gamma_k$ is equal to 1 if switch $k$ is operated. CBs and RECs can be operated directly, however, SECs and LBSs require three switching operations at most (open CB/REC, open/close SEC/LBS, close CB/REC). Therefore, the maximum number of steps is calculated using the following equation:
	{	\color{black}
		{
			\small
			\begin{equation}
			|\Gamma| = \min(\sum_{\forall k \in \Omega_{CB}}\gamma_k^* + 3~ \sum_{\mathclap{\forall k \in \Omega_{Sec} \cup \Omega_{LBS}}}~\gamma_k^*, \bar{\gamma})
			\label{|Gamma|}
			\end{equation}
		}
		
		\noindent where $\gamma_k^*$ is obtained from the optimal topology model, and $\bar{\gamma}$ is the maximum number of switching operations. 

	\subsection{Problem Formulation} 
	
	In this subsection, we formulate SSOP as a MILP model. Since we are only concerned with switches in SSOP, the size of the network can be reduced such that only switchable lines are present. Therefore, we use a network reduction method to ease the modeling procedure and increase the computational efficiency of SSOP, without affecting the solution. The idea is to combine all the buses between switchable lines to form a ``bus block" \cite{B_Chen2018}. Consider the 18-bus distribution network shown in Fig. \ref{Network_Reduction_a}. We first remove all switchable lines and create the subset $\bar{\Omega}_K = \Omega_K\setminus\Omega_{SW}$, which contains non-switchable lines only. Subsequently, Fig. \ref{Network_Reduction_a} is converted to the network shown in Fig. \ref{Network_Reduction_b}. 
	Once all bus blocks are identified, the switchable lines are reinstated, and the reduced network will contain the bus blocks $\Omega_{BL}$ and switchable lines $\Omega_{SW}$. 

	\begin{figure}[htbp!]
		\setlength{\abovecaptionskip}{0pt} 
		\setlength{\belowcaptionskip}{0pt} 
		\centering
		\includegraphics[width=0.48\textwidth]{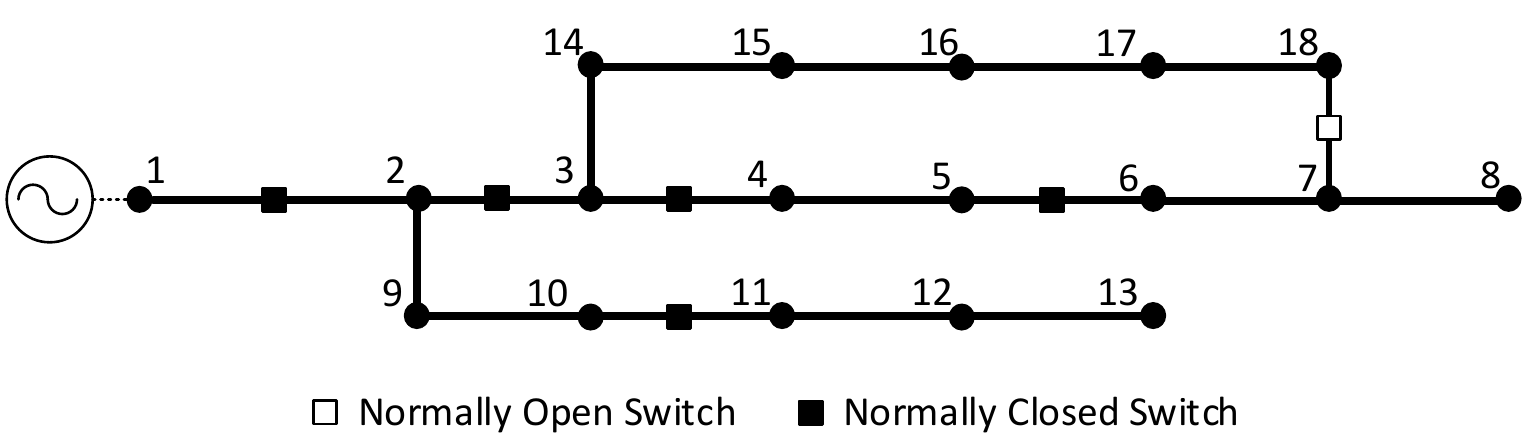}
		\caption{18-bus distribution network with 6 controllable switches.}\label{Network_Reduction_a}
	\end{figure}

	\begin{figure}[htbp!]
		\setlength{\abovecaptionskip}{0pt} 
		\setlength{\belowcaptionskip}{0pt} 
		\centering
		\includegraphics[width=0.48\textwidth]{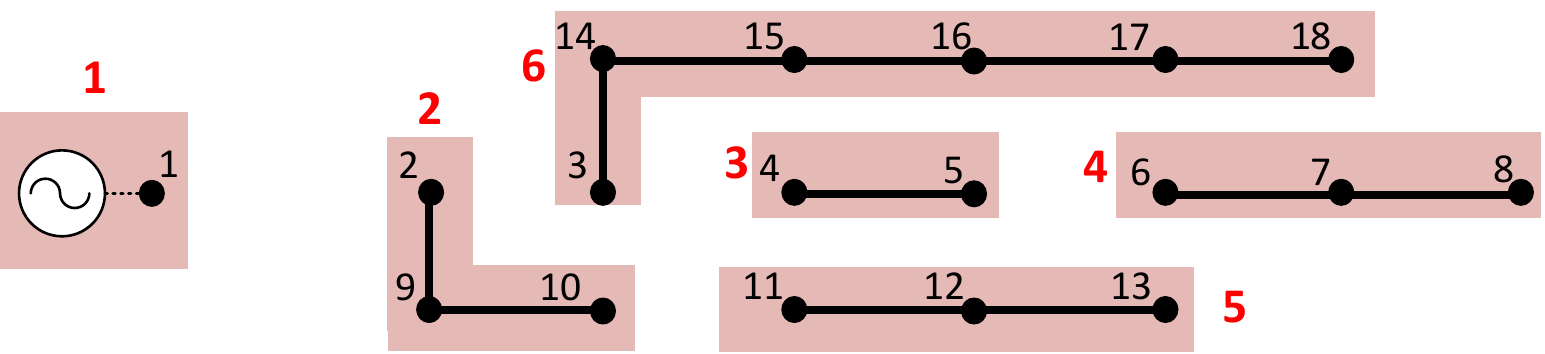}
		\caption{18-bus distribution network with 6 controllable switches removed.}\label{Network_Reduction_b}
	\end{figure}
	
	\noindent
	Next, we formulate the MILP model for SSOP as follows:
	
	\subsubsection{Objective function:}
	
	The objective of the SSOP model is formulated using the following equation:
	
	\vspace{-0.2cm}
	\begin{equation}
	\textrm{max} \sum_{\forall s}\big{(} \sum \limits_{\forall i \in \Omega_{BL}} y_{is} \rho^D_i \sum_{\forall \varphi} \tilde{P}^D_{i\varphi} - \sum_{\mathclap{\forall k \in \Omega_{SW}}} \rho^{SW}_k \gamma_{ks} - \rho^R  \mathcal{O}_s\big{)}
	\label{Sec_Obj}
	\end{equation}
	\noindent
	The objective of the proposed model is to jointly maximize the number of restored loads, minimize the number of switching operations, and minimize the operation time of the switching operations. A penalty price $\rho^R$ is imposed on the total operation time; i.e., penalizing the time it takes to complete the switching operations. The costs, represented by $\rho$, can be considered as weighting factors for the multi-objective equation in (\ref{Sec_Obj}).
	
	\subsubsection{Identify faulted and energized areas}
	
	The variable $x^F_{is}$ is used to identify which bus is in a faulted area and $x^E_{is}$ identifies the bus blocks that are energized. A bus block is considered to be damaged if one line in the bus block is faulted. The following constraints identify the energized and faulted bus blocks: 
	{\small
	\begin{equation}
	x^F_{is} = 1, \forall i \in \Omega_{DB}, s
	\label{Sec_faulted_bus}
	\end{equation}
	\begin{equation}
	x^E_{is} = 1, \forall i \in \Omega_{Sub}, s
	\label{Sec_source_bus}
	\end{equation}
	\begin{equation}
	-(1-u_{ks}) \le x^F_{is} - x^F_{js} \le (1-u_{ks}), \forall k(i,j) \in \Omega_{SW}, s
	\label{Sec_propagate_fault}
	\end{equation}
	\begin{equation}
	-(1-u_{ks}) \le x^E_{is} - x^E_{js} \le (1-u_{ks}), \forall k(i,j) \in \Omega_{SW}, s
	\label{Sec_propagate_source}
	\end{equation}
	\begin{equation}
	y_{is} \le x^E_{is}, \forall i \in \Omega_{BL}, s
	\label{Sec_y_xE}
	\end{equation}
	\begin{equation}
	y_{is} \le 1-x^F_{is}, \forall i \in \Omega_{BL}, s
	\label{Sec_y_xF}
	\end{equation}
}
	\noindent
	Constraint (\ref{Sec_faulted_bus}) sets the value of $x^F_{is}$ to 1 if there is a fault in bus block $i$. Constraint (\ref{Sec_source_bus}) sets $x^E_{is}$ to 1 if bus block $i$ is connected to a substation or generator. If bus $j$ is connected to bus $i$ by switch $k(i,j)$, then the values of $x^E_{is}$ and $x^F_{is}$ should be the same for buses $i$ and $j$, this is enforced in (\ref{Sec_propagate_fault}) and (\ref{Sec_propagate_source}). Therefore, the status (energized/faulted) is propagated around the network based on the connection status of the switches $u_{ks}$. Loads cannot be served if they are not energized (\ref{Sec_y_xE}), and the same applies if the bus is in a faulted area (\ref{Sec_y_xF}). 
	
	\subsubsection{Power operation constraints}
	
	Since the objective of this model is to find the optimal switching sequence, we do not consider detailed distribution system operation constraints. Instead, simplified power flow equations are considered to ensure that a path is available between generators and loads, and that switches operate within their current breaking and making capacities. The constraints are formulated as follows: 
	
	{\small
	\begin{equation}
	0 \le P_{i\varphi s}^G \le \bar{P}^G, \forall i \in \Omega_{BL}, \varphi, s
	\label{SSOP_PG_limit}
	\end{equation}
	\begin{equation}
	-\bar{Q}^G \le Q_{i\varphi s}^G \le \bar{Q}^G, \forall i \in \Omega_{BL}, \varphi, s
	\label{SSOP_QG_limit}
	\end{equation}
	\begin{equation}
	P_{i\varphi s}^G + \sum_{\mathclap{\forall k \in K(.,i)}}P_{k\varphi s} = y_{is} \tilde{P}_{i\varphi}^D+ \sum_{\mathclap{\forall k \in K(i,.)}}P_{k\varphi s}, \forall i \in \Omega_{BL}, \varphi, s
	\label{SSOP_P_flow}
	\end{equation}
	\begin{equation}
	Q_{i\varphi s}^G + \sum_{\mathclap{\forall k \in K(.,i)}}Q_{k\varphi s} = y_{is} \tilde{Q}_{i\varphi}^D+ \sum_{\mathclap{\forall k \in K(i,.)}}Q_{k\varphi s}, \forall i \in \Omega_{BL}, \varphi, s
	\label{SSOP_Q_flow}
	\end{equation}
	\begin{equation}
	P_{k\varphi s}^2+Q_{k\varphi s}^2 \le u_{k t} p_{k \varphi} \bar{S}_k^2, \forall k \in \Omega_{SW}, \varphi, s
	\label{SSOP_S_limit}
	\end{equation}
	\begin{equation}
	\begin{split}
	P_{k \varphi s-1}^2 + &Q_{k \varphi s-1}^2 \le \gamma_{k s} U^n_{i \varphi}\hat{I}_k^2 ~+\\& (1-(u_{k s-1}-u_{k s}))M, \forall k \in \Omega_{SW}, \varphi, s
	\label{breaking_cap}
	\end{split}
	\end{equation}
	\begin{equation}
	\begin{split}
	P_{k \varphi s}^2 + &Q_{k \varphi s}^2 \le \gamma_{k s} U^n_{i \varphi}\check{I}_k^2 ~+\\& (1-(u_{k s}-u_{k s-1}))M, \forall k \in \Omega_{SW}, \varphi, s
	\label{making_cap}
	\end{split}
	\end{equation}
}
	\noindent
	Constraints (\ref{SSOP_PG_limit}) and (\ref{SSOP_QG_limit}) limit the active and reactive power of the generators. The active and reactive power balance equations are modeled in (\ref{SSOP_P_flow}) and (\ref{SSOP_Q_flow}). Constraint (\ref{SSOP_S_limit}) limit the power flow on the lines. The current magnitude on line $k$ equals $S_{k \varphi}/V_{i \varphi}$, where $S_{k \varphi}$ is the apparent power magnitude and $S_{k \varphi}^2 = P_{k \varphi}^2 + Q_{k \varphi}^2$. We estimate the voltage $V_{i \varphi}$ by using the voltage obtained from OTP, which we denote as $V_{i \varphi}^n$. Then, we enforce constraint (\ref{breaking_cap}) so that if a switch is opened ($u_{k s-1}-u_{k s}=1$), the squared current flow $S^2_{k \varphi}/U_{i \varphi}$ through the switch must be less than the squared breaking current $\hat{I}_k^2$ in the previous time step. Similarly, constraint (\ref{making_cap}) states that the squared current flow through the switch must be less than $\check{I}_k^2$ once it is closed. Constraints (\ref{SSOP_S_limit})-(\ref{making_cap}) can be linearized using the circular constraint linearization method \cite{Zhao2019}.

	\subsubsection{Switching constraints}
	
	The next set of constraints are related the status of switches and the operating logic of SECs and LBSs.
	
	\vspace{-0.2cm}
	
	{\small
	\begin{equation}
	u_{k 0} = \Gamma^0_k,  \forall k \in \Omega_{SW}
	\label{Sec_initial_0}
	\end{equation}
	\begin{equation}
	u_{k |\Gamma|} = \Gamma^F_k,  \forall k \in \Omega_{SW}
	\label{Sec_initial_final}
	\end{equation}
	\begin{equation}
	u_{k s} = \Gamma^0_k, \forall k \in \Omega_{FS}, s
	\label{Sec_initial_fuse}
	\end{equation}
	\begin{equation}
	{\gamma}_{k s} \ge u_{k s}-u_{k s-1}, \forall k \in \Omega_{SW}, s, s > 0
	\label{Sec_sw1}
	\end{equation}
	\begin{equation}
	{\gamma}_{k s} \ge u_{k s-1}-u_{k s}, \forall k \in \Omega_{SW}, s, s > 0
	\label{Sec_sw2}
	\end{equation}
	\begin{equation}
	\sum_{\forall k \in \Omega_{SW}} \gamma_{k s} \le 1, \forall s, s > 0
	\label{Sec_one_sw}
	\end{equation}
	\begin{equation}
	\sum_{ \forall k \in \Omega_{SW}} \gamma_{k s} \le \sum_{ \forall k \in \Omega_{SW}} \gamma_{k s-1}, \forall s, s > 1
	\label{Sec_sw_begin}
	\end{equation}
	\begin{equation}
	\gamma_{k s} \le 1 - x^E_{i' s-1}, \forall k(i,j) \in \Omega_{Sec}, i' \in \{i,j\}, s, s > 0
	\label{Sec_section}
	\end{equation}
	\begin{equation}
	\gamma_{k s} \le 2 - x^E_{i s-1} - x^F_{j s-1}, \forall k(i,j) \in \Omega_{LBS}, s, s > 0
	\label{Sec_LBS_1}
	\end{equation}
	\begin{equation}
	\gamma_{k s} \le 2 - x^F_{i s-1} - x^E_{j s-1}, \forall k(i,j) \in \Omega_{LBS}, s, s > 0
	\label{Sec_LBS_2}
	\end{equation}
	}
	\noindent
	Constraints (\ref{Sec_initial_0}) and (\ref{Sec_initial_final}) define the initial and final status of each switch, respectively. The final status of each switch, $\Gamma^F_k$, is determined by solving OTP. Constraint (\ref{Sec_initial_fuse}) indicates that the status of a line with a fuse does not change. Melted fuses are replaced manually by the repair crews. Constraints (\ref{Sec_sw1}) and (\ref{Sec_sw2}) are used to calculate the value of $\gamma_{ks}$, which equals 1 if switch $k$ is opened or closed in step $s$. There can only be one switching operation in each step, as enforced by (\ref{Sec_one_sw}). Constraint (\ref{Sec_sw_begin}) ensures that the switching operations are not delayed to the last steps. SECs cannot operate if they are energized, which is realized by constraint (\ref{Sec_section}). Constraints (\ref{Sec_LBS_1}) and (\ref{Sec_LBS_2}) ensure that an LBS can only be operated if it is not in an energized and faulted area at the same time, i.e., fault current is not running through the LBS.

	\subsubsection{Manual switches}
	
	Operating a manual switch when it is energized can be life-threatening. Distribution system operators must ensure that manual switches are de-energized before specialized field crews operate them. Coordinating remotely controllable switches and manual switches can be challenging due to the difference in operation times \cite{BChen_2019}. Operating a remotely controllable switch requires a few seconds, while a manually operated switch takes several minutes or hours. In this paper, we model the operation of manual switches by incorporating the Vehicle Routing Problem (VRP) \cite{Laporte_2009} in SSOP. The variable $x_{klc}$ represents the path a crew takes, if crew $c$ travels from switch $k$ to switch $l$, then $x_{klc}=1$. The constraints are formulated as follows:
	
	{\small
	\begin{equation}
	\sum_{\forall k \in \hat{\Omega}_{MS}}\sum_{\forall c} x_{k l c} = \sum_{\forall s}\gamma_{l s}, \forall l \in {\Omega}_{MS}
	\label{visit_MS}
	\end{equation}
	\begin{equation}
	\mathop \sum \limits_{\forall k \in \hat{\Omega}_{MS}}{x_{0 k c}} = 1, \forall c
	\label{start_from_depot_MS}
	\end{equation}
	\begin{equation}
	\mathop \sum \limits_{ \forall k \in \hat{\Omega}_{MS}} {x_{k 0 c}} = 1, \forall c
	\label{back_to_depot_MS}
	\end{equation}
	\begin{equation}
	\mathop \sum_{{\forall l \in \hat{\Omega}_{MS}\backslash \left\{ k \right\}}}
	{x_{k l c}} - \mathop \sum_{{{ \forall l \in \hat{\Omega}_{MS}\backslash \left\{ k \right\}}}} {x_{l k c}} = 0, \forall c,k \in {\Omega}_{MS}
	\label{path_flow_MS}
	\end{equation}
	\begin{equation}
	\begin{split}
	\alpha_{k} + w_k + \mathcal{T}^S_k + & t{r_{k l}} - \left( {1 - \sum_{\forall c}{x_{k l c}}} \right)M \\ \le \alpha_{l},~
	& \forall k \in \hat{\Omega}_{MS},l \in {\Omega}_{MS}, k \neq l
	\end{split}
	\label{Arrival_const_MS}
	\end{equation}
	\begin{equation}
	\begin{split}
	\alpha_{k} + w_k + \mathcal{T}^S_k + &t{r_{k l}} + \left( {1 - \sum_{\forall c}{x_{k l c}}} \right)M \\ \ge \alpha_{l},~
	& \forall k \in \hat{\Omega}_{MS},l \in {\Omega}_{MS}, k \neq l
	\end{split}
	\label{Arrival_const_MS_2}
	\end{equation}
	\begin{equation}
	0 \le w_k \le \bar{w}, \forall k \in \Omega_{MS}
	\label{max_wait}
	\end{equation}
	\begin{equation}
	\mathcal{O}_s \ge \alpha_k + w_k + \mathcal{T}^S_k - M (1-\gamma_{k s}), \forall k \in {\Omega}_{MS}, s
	\label{operation_MS}
	\end{equation}
	\begin{equation}
	\alpha_k + w_k\ge \mathcal{O}_{s-1} - M (1-\gamma_{k s}), \forall k \in \Omega_{MS}, s
	\label{AT>OP}
	\end{equation}
	\begin{equation}
	\mathcal{O}_s \ge \mathcal{O}_{s-1} ~~+~~ \sum_{\mathclap{\forall k \in \Omega_{SW}\backslash\Omega_{MS}}} ~~\mathcal{T}^S_k \gamma_{k s}, \forall s
	\label{operation_prev}
	\end{equation}}
	\noindent
	Constraint (\ref{visit_MS}) states that a crew visits a manual switch if it is scheduled to be operated. The set $\hat{\Omega}_{MS}$ is the union of $ \Omega_{MS}$ and $\{0\}$, where $\{0\}$ represents the depot (starting location of the crews). 
	Constraints (\ref{start_from_depot_MS})--(\ref{back_to_depot_MS}) define the starting and ending locations for the crews. Equation (\ref{path_flow_MS}) represents the path-flow constraint for the routing problem. The arrival time is calculated in (\ref{Arrival_const_MS}) and (\ref{Arrival_const_MS_2}), where $\alpha_k + w_k + \mathcal{T}^S_k + tr_{kl} = \alpha_l$ if a crew travels from $k$ to $l$. The waiting time $w_k$ represents the time between arrival and start of switching operation, which is constrained by (\ref{max_wait}). We assume the maximum wait time is 30 minutes in this study. In order to calculate the time elapsed between the switching operations, we define the variable $\mathcal{O}_s$. For manual switches, $\mathcal{O}_s$ equals the arrival time plus the operating time of a manual switch and waiting time, as defined in (\ref{operation_MS}), where the constraint is applied only if switch $k$ is operated in step $s$. If switch $k$ is to be operated in step $s$, then the arrival time added to the waiting time at $k$ should be higher or equal to $\mathcal{O}_{s-1}$, which is represented in (\ref{AT>OP}). Constraint (\ref{operation_prev}) calculates the elapsed time by adding the operation time of the automatic switches.
	
	\vspace{-0.25cm}
	\subsection{Fault Repair}
	
	After performing the switching operations, we dispatch the repair crews to the faulted lines in the system. The repair crew routing problem is solved separately from OTP and SSOP due to the difference in time scale, however, we still consider distribution system constraints when dispatching crews. RCRP is modeled by coupling constraints from OTP and VRP. The problem can be defined by a complete undirected graph $\mathcal{G}$ with nodes ($\mathcal{N}$) and edges ($E$). In previous work \cite{Arif2018b},  VRP was combined with distribution system operation constraints, creating the distribution system repair and restoration problem (DSRRP). In this paper, we leverage the bus blocks concept to design the graph $\mathcal{G}$. Instead of routing the crews to each damaged components, we route the crews to bus blocks so that the nodes are equal to the set of damaged bus blocks $\Omega_{DB}$. Crews that travel to bus blocks are then assigned to the damaged components inside the bus blocks. The idea is that the travel time between components inside a bus block is small, compared to the repair times and the travel times between the bus blocks, and therefore can be neglected. The crew routing problem is depicted by Fig. \ref{VRP_BusBlocks}. {\color{black} A description for the mathematical model is given below: 
		
		\[
		\text{\textbf{min} \{load shedding costs + travel costs\} 
		}
		\]
		\[
		\text{\textbf{subject to}} 
		\begin{cases}
		\text{Routing to bus blocks and assignment}
		\\
		\text{Arrival and repair times} 
		\\
		\text{Distribution system constraints}
		\end{cases}
		\]
		\noindent
		The mathematical model for RCRP can be found in Appendix B.} 
	Once crews repair a section of the network, we solve OTP and SSOP again to update the topology of the network. 
	
	\begin{figure}[h!]
		\setlength{\abovecaptionskip}{0pt} 
		\setlength{\belowcaptionskip}{0pt} 
		\centering
		\includegraphics[width=0.48\textwidth]{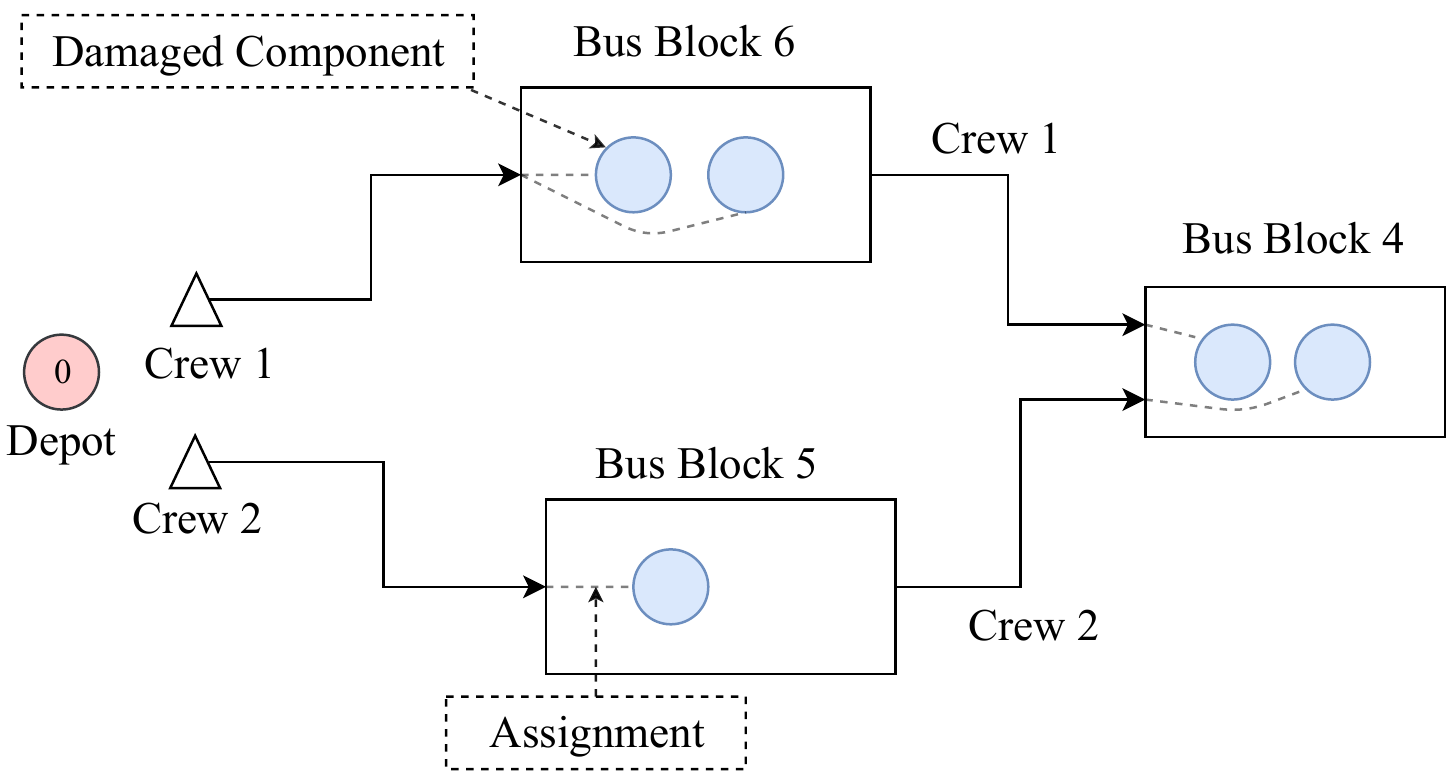}
		\caption{Vehicle routing problem converted from Fig. \ref{Network_Reduction_b} with 5 damaged lines.}\label{VRP_BusBlocks}
	\end{figure}
	
	\vspace{-0.2cm}

	\section{Simulation and Results}\label{sec:7}

	Modified versions of the IEEE 123-bus distribution system and the IEEE 8500-bus system are used as test cases in this paper. The operation times of manual and remotely-controllable switches are set to 15 and 1 minutes, respectively. We assume the breaking and making current capacities are the same. LBSs are rated at 500 A. CBs and RECs are rated to interrupt fault currents. 
	SECs cannot make or break currents, therefore, they are rated at 0 A. Also, we assume the maximum number of switching operations is 25. The simulated problems are modeled in AMPL and solved using GUROBI 9.0 on a PC with Intel Core i7-8550U 1.8 GHz CPU and 16 GB RAM. {\color{black} Five test cases are simulated in this section. The first four test cases are conducted on the IEEE 123-bus distribution system, and the fifth test is conducted on the IEEE 8500-bus system.}
	
	

	\subsection{Test Case I}
	
	The modified IEEE 123-bus network contains 6 CBs, 11 RECs, 4 LBSs, 17 SECs, and 14 Fuses. The initial status of each switch is shown in Fig. \ref{IEEE123_Modified}. 
	SECs 54-94, 60-160, and 78-80 are assumed to be manual switches (must be operated by a crew), while all CBs, RECs, and LBSs are remotely controllable. The power supplied by the substations are limited to 2 MW and 1 {\color{black}Mvar} per-phase. The network reduction algorithm is used to reduce the system, the reduced network has 51 bus blocks.
	
	\begin{figure}[h!]
		\setlength{\abovecaptionskip}{0pt} 
		\setlength{\belowcaptionskip}{0pt} 
		\centering
		\includegraphics[width=0.49\textwidth]{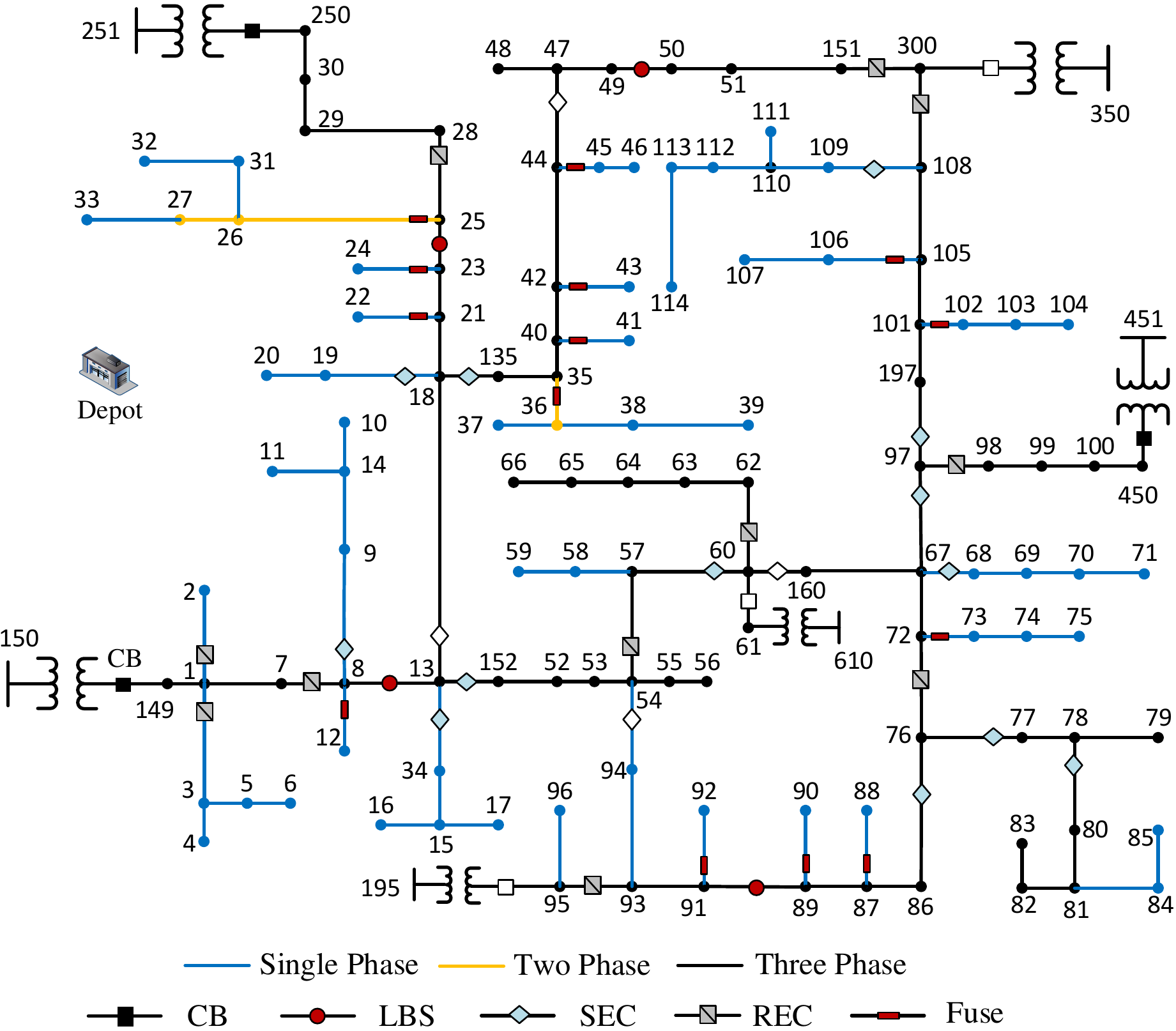}
		\caption{Modified IEEE 123-bus distribution system. A shaded switch indicates that the switch is closed.}\label{IEEE123_Modified}
	\end{figure}

	A permanent fault is assumed to have occurred on line 18-21, and REC 25-28 was opened to clear the fault. To test the operation of the LBSs, we simulate the problem with the LBSs rated at 500 A, and then decrease the rating to 50 A. OTP is first solved to obtain the optimal final state of each switch. SSOP is then solved to find the optimal sequence of operations to reach the desired topology obtained from OTP. The solutions are shown in Table \ref{Results_Test_I}. The computation time is 0.2 s for OTP, and 3.23 s for SSOP. OTP finds that LBS 23-25 and SEC 18-135 must be opened, while REC 25-28 and SEC 44-47 should be closed. However, it is not possible to directly operate these switches due to their characteristics. If the LBSs are rated at 500 A, the switching sequence starts by opening SEC 18-135 to isolate buses 35--46 from the fault. The next step is to open LBS 49-50 in order to close SEC 44-47 in the following step. In the fourth step, LBS 49-50 is closed and buses 35--46 are energized. In step 5, LBS 23-25 is opened, which isolates  buses 25--33 from the fault on line 18-21. Finally, buses 25--33 are energized by closing REC 25-28. After changing the rating of the LBSs to 50 A, the sequence remains the same except for the operation of LBS 49-50. The LBS cannot be operated due to its low current capacity. Instead of operating LBS 49-50, REC 108-300 is opened and closed in steps 2 and 4, respectively. On the other hand, LBS 23-25 can be opened as buses 23 and 25 are not energized.
	
	
	\begin{table}[htbp]
		\centering
		\small
		\caption{Switching Operations For Test Case I}
		\vspace{-0.2cm}
		\begin{threeparttable}
			\begin{tabular}{|p{20mm}| p{60mm}|}
				\hline
				\rule{0pt}{2.2ex}Stage & Switching Operations \\
				\hline
				\rule{0pt}{2.2ex}Fault Clearance & $\uparrow$ REC 25-28\\
				\hline
				\rule{0pt}{2.2ex}OTP & $\uparrow$ SEC 18-135, $\downarrow$ SEC 44-47, $\uparrow$ LBS 23-25, \newline $\downarrow$ REC 25-28\\
				\hline
				\rule{0pt}{2.2ex}SSOP \newline LBS: 500 A & $\uparrow$ SEC 18-135, $\uparrow$ LBS 49-50, $\downarrow$ SEC 44-47, \newline $\downarrow$ LBS 49-50, $\uparrow$ LBS 23-25, $\downarrow$ REC 25-28  \\
				\hline
				\rule{0pt}{2.2ex}SSOP \newline LBS: 50 A  & $\uparrow$ SEC 18-135, $\uparrow$ REC 108-300, $\downarrow$ SEC 44-47, \newline  $\downarrow$ REC 108-300, $\uparrow$ LBS 23-25, $\downarrow$ REC 25-28  \\
				\hline
			\end{tabular}%
			\label{Results_Test_I}%
			{$\uparrow$: open switch, $\downarrow$: close switch.
			}
		\end{threeparttable}
	\end{table}%
	\vspace{-0.4cm}
	{\color{black}
		\subsection{Test Case II}
		
		In the second test case, lines 28--29, 51--151, 99--100, and 105--108 are assumed to be damaged. The initial state of the network after the damage is given in Fig. \ref{IEEE123_Modified_new_case}, where the shaded portion indicates energized lines. The purpose of this test case is to compare the proposed method with the common approach in the literature, which assumes a uniform type of switches without operational constraints (i.e., all switches have the capabilities of CBs/RECs) \cite{Carvalho2007,Li2014,B_Chen_multistep,B_Chen2018}. 
		
		\begin{figure}[h!]
			\setlength{\abovecaptionskip}{0pt} 
			\setlength{\belowcaptionskip}{0pt} 
			\centering
			\includegraphics[width=0.49\textwidth]{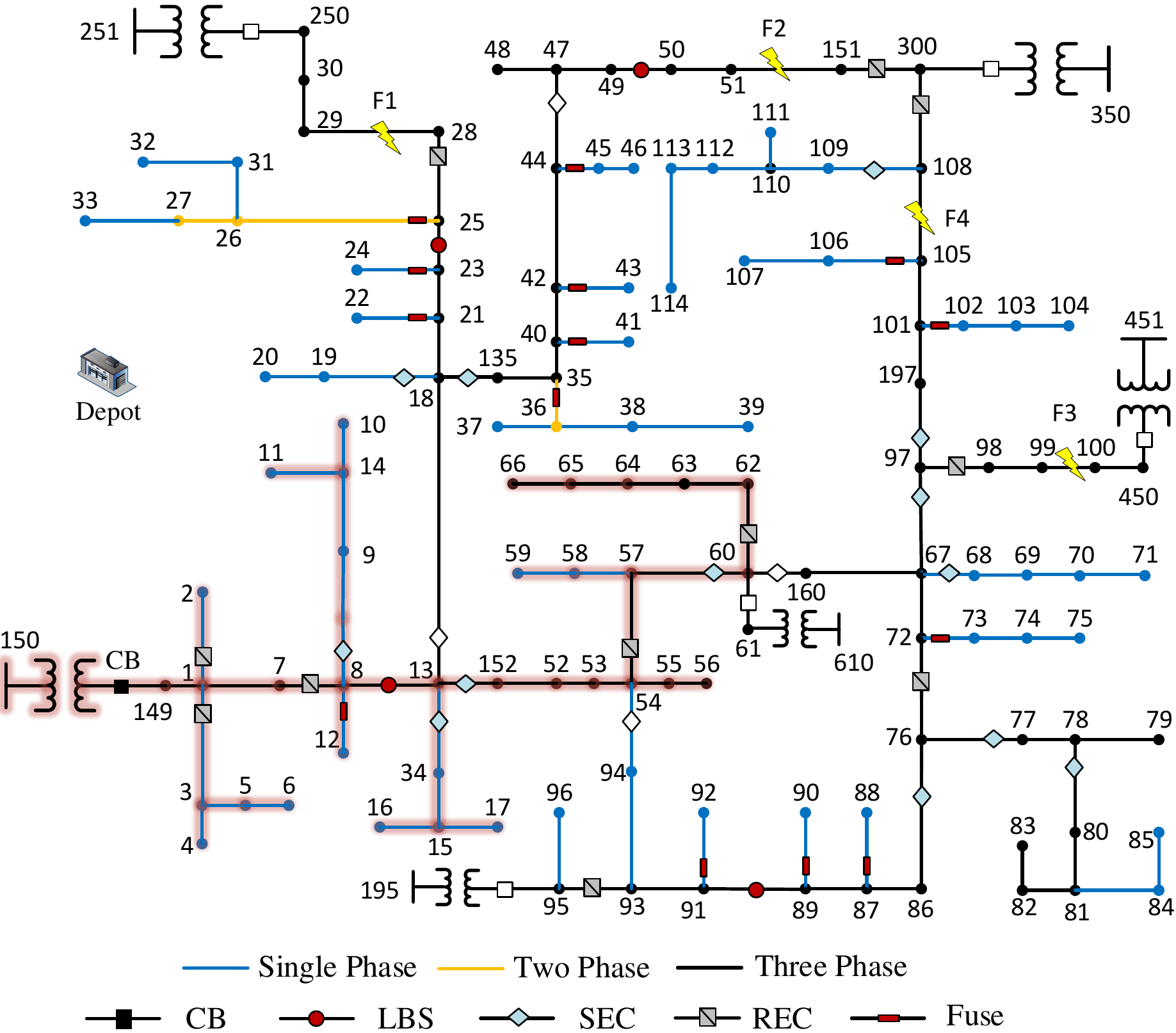}
			\caption{Initial state of the IEEE 123-bus network in after 4 lines are damaged.}\label{IEEE123_Modified_new_case}
		\end{figure}

		The sequence of switching operations are shown in Table \ref{Sol_switching_table_case2_new}, where invalid operations are highlighted in bold. With uniform switches, SEC 67-97 is opened to isolate F2--F4 from the substation 195. The CB at substation 195 is then closed to supply loads 67--96. Next, REC 25-28 is opened to isolate F1 and SEC 13-18 is closed to restore loads 18--27 and 31--33. However, closing SEC 13-18 at this stage is not possible in practice, as bus 13 is energized and SECs can only operate under no-load condition. The LBS 49-50 is then opened and SEC 44-47 is closed to restore loads 35--46. Again, this last SEC operation is invalid since bus 44 is energized. Neglecting the capabilities of different switches leads to switching steps that are inapplicable. 
		
		Next, we show the correct sequence of switching operations using the proposed method. The first two operations are the same, where SEC 67-97 is opened and CB 95-195 is closed. SEC 44-47 is then closed and both REC 25-28 and LBS 49-50 are opened. Subsequently, loads 18--27, 32--33, and 35--49 can receive energy from substation 150 if SEC 13-18 is closed. However, LBS 8-13 must be opened first before closing SEC 13-18 to de-energize bus 13, and LBS 8-13 is then closed in the final step. 
		The results show the importance of including device-specific constraints to achieve solutions that can be applied in practice.   
	}
	
	\vspace{-0.2cm}
	\begin{table}[htbp]
		\centering
		\small
		\caption{Switching Operations For Test Case II}
		\vspace{-0.2cm}
		\begin{threeparttable}
			\begin{tabular}{|p{12mm}| p{55mm}| p{10mm}|}
				\hline
				\rule{0pt}{2.2ex}Method & Switching Operations & Comp. Time \\
				\hline
				\rule{0pt}{2.2ex}Uniform switches &   $\uparrow$ SEC 67-97, $\downarrow$ CB 95-195, $\uparrow$ REC 25-28, $\downarrow$ \textbf{SEC 13-18}, $\uparrow$ LBS 49-50, $\downarrow$ \textbf{SEC 44-47}   & ~~11 s\\
				\hline
				\rule{0pt}{2.2ex}Proposed Method &   $\uparrow$ SEC 67-97, $\downarrow$ CB 95-195, $\downarrow$ SEC 44-47, $\uparrow$ REC 25-28, $\uparrow$ LBS 49-50, $\uparrow$ LBS 8-13, $\downarrow$ SEC 13-18, $\downarrow$ LBS 8-13    & ~~19 s \\
				\hline
			\end{tabular}%
			{$\uparrow$: open switch, $\downarrow$: close switch.}
		\end{threeparttable}
		\label{Sol_switching_table_case2_new}%
	\end{table}%

	
	\subsection{Test Case III}
	
	In the third test case, we simulate 7 damaged lines on the IEEE 123-bus system and solve the service restoration problem using the process shown in Fig. \ref{Flowchart}. The simulated damage and initial status of each switch ($\Gamma^0_k$) are shown in Fig. \ref{IEEE123_Modified_2}. The numbers of operation crews (for operating manual switches) and line crews are assumed to be 2 and 3, respectively. Travel times are estimated using the Euclidean distances, we scale the travel times so that they range between 5 to 30 minutes. The repair times, which are determined by the damage assessors, are assumed to be between 30 minutes to 3 hours.

	\begin{figure}[h!]
		\setlength{\abovecaptionskip}{0pt} 
		\setlength{\belowcaptionskip}{0pt} 
		\centering
		\includegraphics[width=0.49\textwidth]{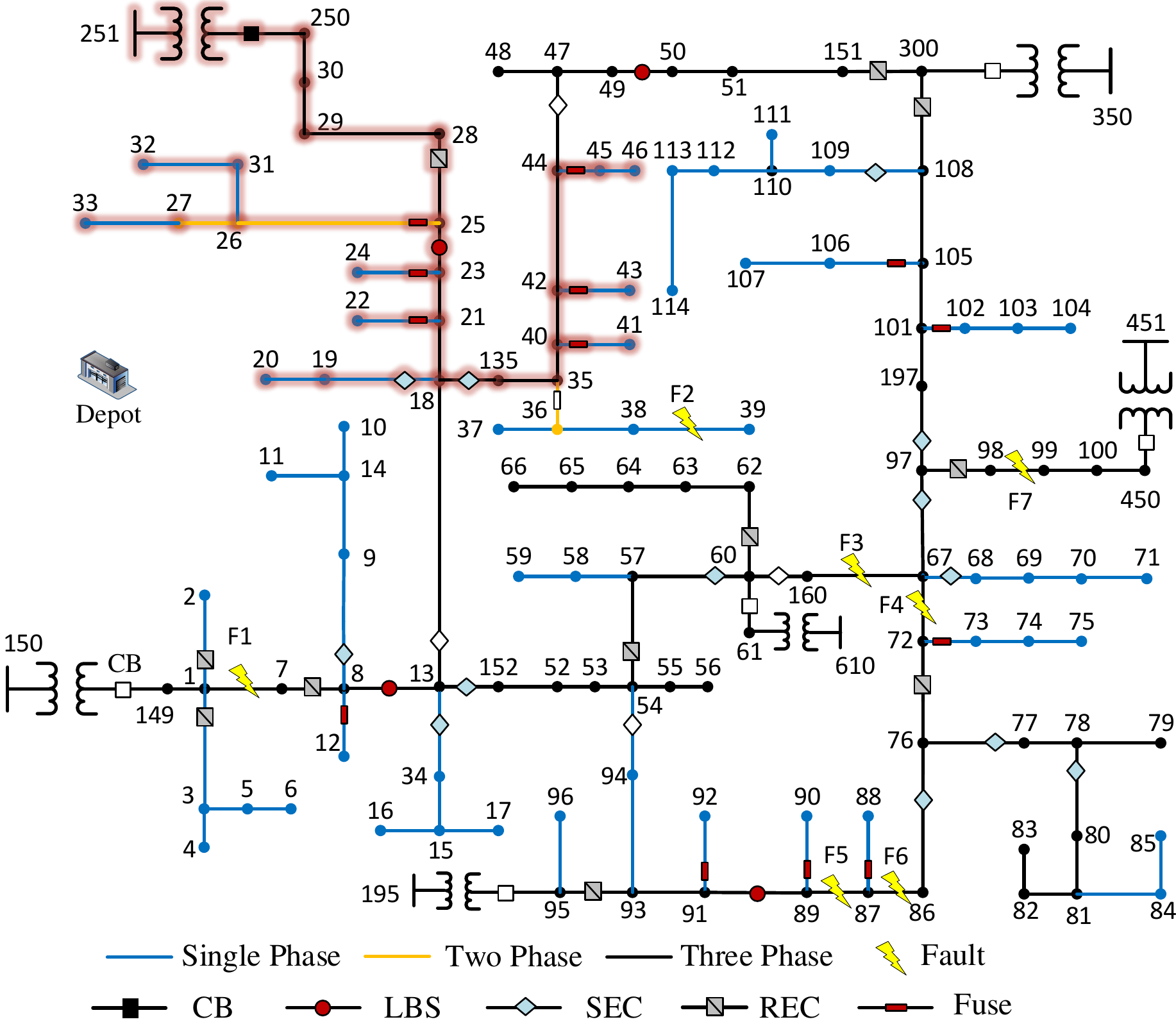}
		\caption{Initial state of the IEEE 123-bus network after 7 lines are damaged.}\label{IEEE123_Modified_2}
	\end{figure}
	
	There are 5 damaged bus blocks in the simulated test case. For example, the bus block containing buses 86--89 is damaged by F5 and F6. OTP is initially solved to obtain $\Gamma^F_k$, which represents the target topology before conducting any repairs. SSOP is then solved to obtain the sequence of switching operations. RCRP is solved to route the repair crews. Once a section (bus block) in the network is repaired, we solve OTP and SSOP again to update the topology. The sequential operations of the switches, before and after the repairs, are presented in Table \ref{Sol_switching_table_case2},  {\color{black} while the change in number of served loads is shown in Fig. \ref{Restored_Load}}. The routing solution and the topology before the repairs are shown in Fig. \ref{IEEE123_Modified_3}. The first step is to open SEC 97-197 to isolate substation 350 from F3--F7, and then CB 300-350 is closed, which allows substation 350 to supply the loads at buses 47--51 and 101--114. Next, REC 7-8 is opened to isolate F1. SEC 13-18 cannot be closed since bus 18 is energized, therefore, LBS 23-25 is first opened and then closed after closing SEC 13-18. By closing SEC 13-18, a path is provided for substation 251 to supply some of the loads, as shown in Fig. \ref{IEEE123_Modified_3}. LBS 89-91 is then opened to isolate F5 and F6 from substation 195, which supplies buses 91--96 after closing CB 95-195.

	\begin{table}[htbp]
		\centering
		\small
		\caption{Switching Operations For Test Case III}
		\vspace{-0.2cm}
		\begin{threeparttable}
			\begin{tabular}{|p{10mm}| p{55mm}| c|}
				\hline
				\rule{0pt}{2.2ex}Repair & Switching Operations & Comp. Time \\
				\hline
				\rule{0pt}{2.2ex}-- &   $\uparrow$ SEC 97-197, $\downarrow$ CB 300-350, $\uparrow$ REC 7-8, $\uparrow$ LBS 23-25, $\downarrow$ SEC 13-18, $\downarrow$ LBS 23-25, $\uparrow$ LBS 89-91, $\downarrow$ CB 95-195   & 8 s\\
				\hline
				\rule{0pt}{2.2ex}F2 & Replace Fuse 35-36 & NA \\
				\hline
				\rule{0pt}{2.2ex}F3, F4 &   $\uparrow$ SEC 76-86, $\uparrow$ SEC 67-97, $\uparrow$ REC 54-57, $\downarrow$ SEC 60-160, $\downarrow$ REC 54-57    & 3 s \\
				\hline
				\rule{0pt}{2.2ex}F1 &   $\downarrow$ REC 7-8    & 0.2 s \\
				\hline
				\rule{0pt}{2.2ex}F5, F6, F7 &   $\downarrow$ LBS 89-91, $\uparrow$ REC 108-300, \newline $\downarrow$ SEC 97-197, $\downarrow$ REC 108-300    & 0.45 s \\
				\hline
			\end{tabular}%
			{$\uparrow$: open switch, $\downarrow$: close switch.}
		\end{threeparttable}
		\label{Sol_switching_table_case2}%
	\end{table}%

	\begin{figure}[h!]
		\setlength{\abovecaptionskip}{0pt} 
		\setlength{\belowcaptionskip}{0pt} 
		\centering
		\includegraphics[width=0.49\textwidth]{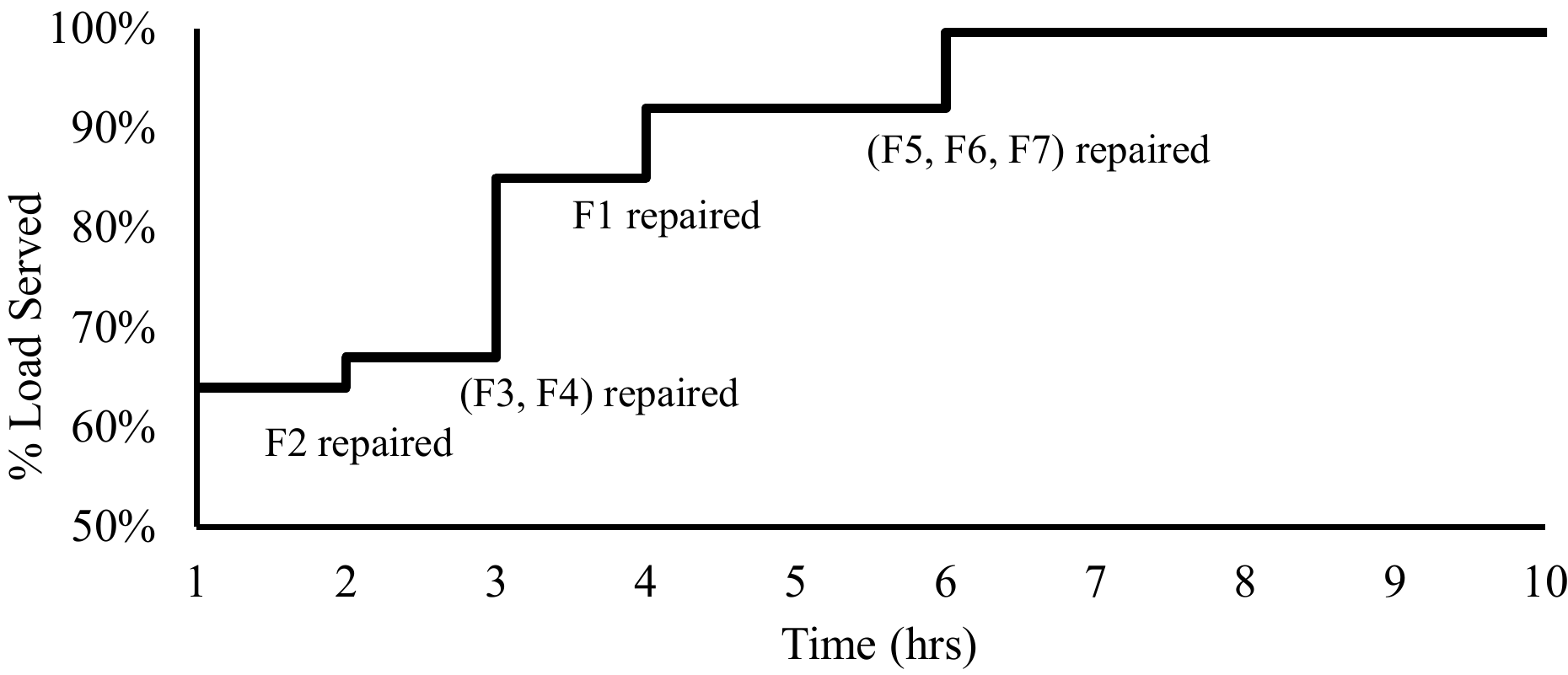}
		\caption{{\color{black} Change in percentage of restored load with time for test case III.}}\label{Restored_Load}
	\end{figure}
	
	\begin{figure}[h!]
		\setlength{\abovecaptionskip}{0pt} 
		\setlength{\belowcaptionskip}{0pt} 
		\centering
		\includegraphics[width=0.49\textwidth]{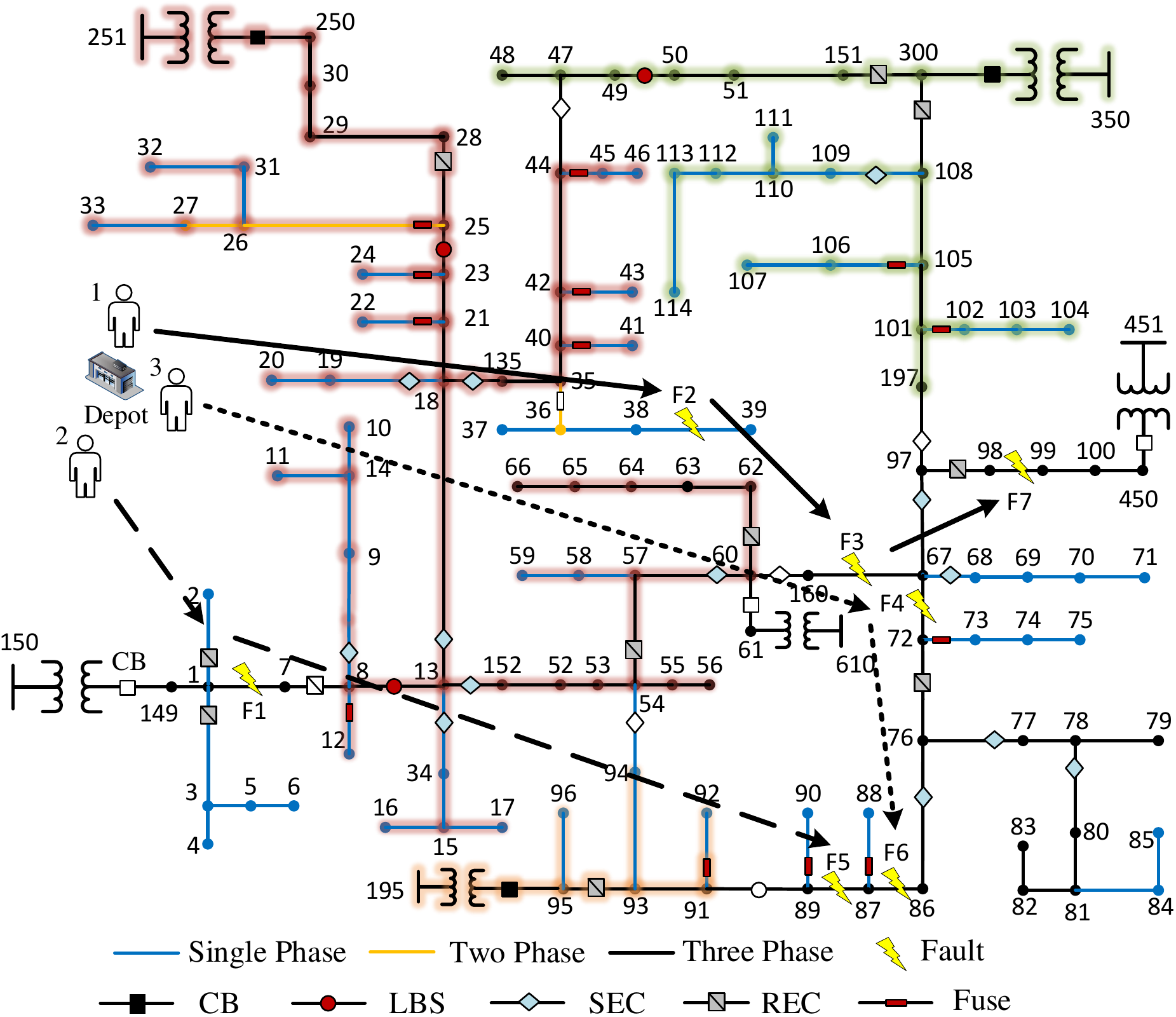}
		\caption{First sequential switching operation and crew routing for test case II.}\label{IEEE123_Modified_3}
	\end{figure}

	After crew 1 repairs F2, the crew replaces fuse 35-36 and no switching operation is required. The next switching operation occurs after crews 1 and 2 repair F3 and F4. Without the two faults, we are able to serve buses 67--85. To achieve that, SECs 76-86 and 67-97 are opened to isolate faults F5--F7. Before operating the manual switch SEC 60-160, REC 54-57 must be opened to de-energize bus 60. REC 54-57 is closed after operating SEC 60-160, which provides a path for substation 251 to supply buses 67--85. {\color{black} At this point, around 85\% of the loads are served (see Fig. \ref{Restored_Load})}. REC 7-8 is closed after clearing F1, subsequently, all loads on the left side of the network can be served. Once all lines are repaired, LBS 89-91 is closed and substation 195 restores buses 86--90. The next step is to serve buses 98--100. REC 108-300 is opened to de-energize bus 197, and SEC 97-197 is then closed. Finally, REC 108-300 is closed and all loads are restored. 


	For the routing solution, we compare the route obtained using RCRP to DSRRP from \cite{Arif2018b}. The proposed crew routing method considers less routing variables and a simplified distribution system operation model. By using network reduction, the number of buses and routing variables are reduced by more than half, as shown in Table \ref{Sol_routing_table}. The methods achieved the same solution, where the total energy served is 80,390 kWh. However, the computation time for RCRP is 75 seconds, which is significantly less than DSRRP (38 minutes).

	\begin{table}[htbp]
		\centering
		\small
		\caption{Performance of Repair Crew Routing for Test Case II}
		\vspace{-0.2cm}
		\begin{threeparttable}
			\begin{tabular}{ccccc}
				\hline
				\rule{0pt}{2.2ex}~~~~Method~~~~ & Buses & Routing Var. & Comp. Time & ~~ES (kWh)~~\\
				\hline
				\rule{0pt}{2.2ex}DSRRP \cite{Arif2018b} & 129   & 192   & \textbf{38 min} & 80,390 \\
				\rule{0pt}{2.2ex}RCRP & 51    & 75    & \textbf{75 s} & 80,390 \\
				\hline
			\end{tabular}%
			{Routing Var.: number of routing variables $\grave{x}_{ijc}$, ES: energy served.}
		\end{threeparttable}
		\label{Sol_routing_table}%
	\end{table}%
	{\color{black}

		\vspace{-0.2cm}
		\subsection{Test Case IV}
		
		In this test case, we modify the IEEE 123-bus distribution system by including five 800 kW dispatchable distributed generators (DGs) and demonstrate how microgrids can be formed around the DGs. Each DG is equipped with a CB, and we assume the CBs are initially open. Moreover, we compare SSOP with two benchmark methods, which are adapted from \cite{Lopez2018} and \cite{B_Chen_multistep}. The modified system, with its initial status after four lines are damaged, is shown in Fig. \ref{MG_test_1}. For this test case, we assume that only substations 150 and 251 can supply power. The switching operations for SSOP, benchmark method A \cite{B_Chen_multistep}, and benchmark method B \cite{Lopez2018} are shown in Table \ref{Sol_switching_table_case4}.
		The first two switching actions in SSOP is to open SEC 18-135 and 108-300 to isolate faults F1 and F3 from DG 48. SEC 44-47 is then closed and the DG at bus 48 is connected to serve the loads on buses 37--51, creating a microgrid in the area, as shown in Fig. \ref{MG_test_2}. Next, REC 54-57 is opened to isolate F2 and the DG at bus 62 is connected to serve loads 57-66. SEC 97-197, SEC 76-77, and LBS 89-91 are opened to isolate faults F3, F4, and F5, respectivley. Before connecting DG 99, SEC 76-86 is closed since it can only operate under no-load condition, and then CB 99 is closed to create another microgrid. SEC 13-152 is opened to isolate F2 and REC 7-8 is closed in order to connect buses 8-17 and 34 to substation 150. SEC 78-80 is opened and CB 83 is closed to serve load 80-85. Finally, LBS 23-25 is opened to isolate F1 and REC 25-28 is closed to serve 25-33. The final circuit is shown in Fig. \ref{MG_test_2}. For the benchmark methods, both achieve the same switching solution. The differences between the benchmark methods and SSOP is given in bold in Table \ref{Sol_switching_table_case4}, where CB 48 and CB 99 are closed before closing SEC 44-47 and SEC 76-87, respectively. Notice that after closing CBs 48 and 99, buses 47 and 76 will be energized, therefore, we cannot operate SECs 44-47 and 76-87 since they do not have current making capabilities. Compared to SSOP, the benchmark methods do not always provide feasible sequential switching operations. The computation time is 30 s for SSOP, 18 s for method A, and 67 s method B. Method B has a higher computation time due to a more complex optimization model. SSOP is marginally slower than method A since we consider the interactions between the switches and their characteristics.
		
		\begin{figure}[h!]
			\setlength{\abovecaptionskip}{0pt} 
			\setlength{\belowcaptionskip}{0pt} 
			\centering
			\includegraphics[width=0.49\textwidth]{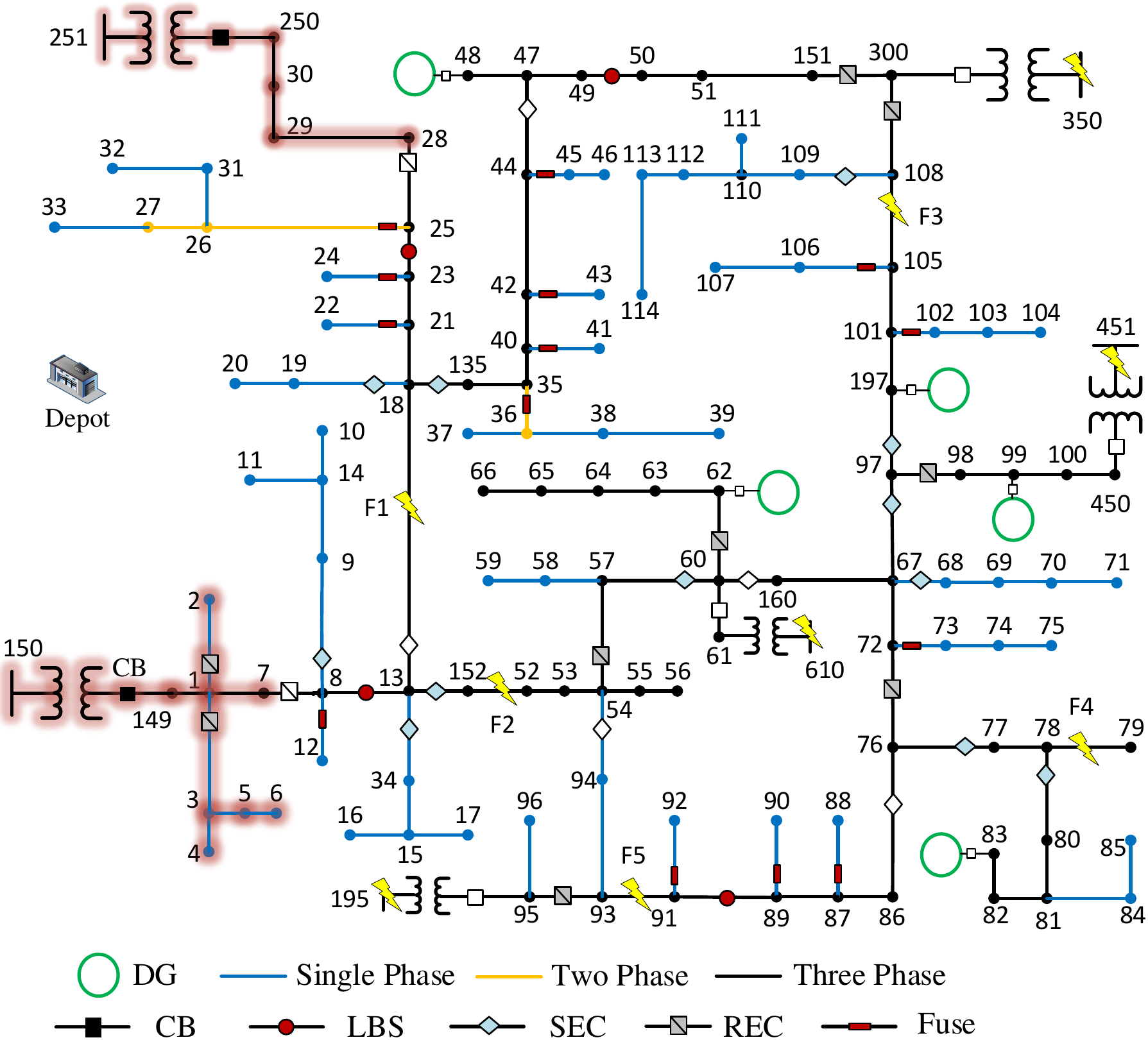}
			\caption{Initial state of the modified IEEE 123-bus system with five damaged lines in test case IV.}\label{MG_test_1}
		\end{figure}

		\begin{table}[htbp]
			\centering
			\small
			\caption{Switching Operations For Test Case IV}
			\vspace{-0.2cm}
			\begin{threeparttable}
				\begin{tabular}{|p{13mm}| p{60mm}| p{8mm}|}
					\hline
					\rule{0pt}{2.2ex}Method & Switching Operations & Comp. Time \\
					\hline
					\rule{0pt}{2.2ex}SSOP &   $\uparrow$ SEC 18-135,
					$\uparrow$ SEC 108-300,
					$\downarrow$ SEC 44-47,
					$\downarrow$ CB 48,
					$\uparrow$ REC 54-57,
					$\downarrow$ CB 62,
					$\uparrow$ SEC 97-197,
					$\uparrow$ SEC 76-77,
					$\uparrow$ LBS 89-91,
					$\downarrow$ SEC 76-86,
					$\downarrow$ CB 99,
					$\uparrow$ SEC 13-152,
					$\downarrow$ REC 7-8,
					$\uparrow$ SEC 78-80,
					$\downarrow$ CB 83,
					$\uparrow$ LBS 23-25,
					$\downarrow$ REC 25-28   & ~~30 s\\
					\hline
					\rule{0pt}{2.2ex}Method A \cite{B_Chen_multistep} &   $\uparrow$ SEC 18-135,
					$\uparrow$ SEC 108-300,
					\textbf{$\downarrow$ CB 48},
					\textbf{$\downarrow$ SEC 44-47},
					$\uparrow$ REC 54-57,
					$\downarrow$ CB 62,
					$\uparrow$ SEC 97-197,
					$\uparrow$ SEC 76-77,
					$\uparrow$ LBS 89-91,
					\textbf{$\downarrow$ CB 99},
					\textbf{$\downarrow$ SEC 76-86},
					$\uparrow$ SEC 13-152,
					$\downarrow$ REC 7-8,
					$\uparrow$ SEC 78-80,
					$\downarrow$ CB 83,
					$\uparrow$ LBS 23-25,
					$\downarrow$ REC 25-28    & ~~18 s \\
					\hline
					\rule{0pt}{2.2ex}Method B \cite{Lopez2018} &   
					$\uparrow$ SEC 18-135,
					$\uparrow$ SEC 108-300,
					\textbf{$\downarrow$ CB 48},
					\textbf{$\downarrow$ SEC 44-47},
					$\uparrow$ REC 54-57,
					$\downarrow$ CB 62,
					$\uparrow$ SEC 97-197,
					$\uparrow$ SEC 76-77,
					$\uparrow$ LBS 89-91,
					\textbf{$\downarrow$ CB 99},
					\textbf{$\downarrow$ SEC 76-86},
					$\uparrow$ SEC 13-152,
					$\downarrow$ REC 7-8,
					$\uparrow$ SEC 78-80,
					$\downarrow$ CB 83,
					$\uparrow$ LBS 23-25,
					$\downarrow$ REC 25-28    & ~~67 s \\
					\hline
				\end{tabular}%
				{$\uparrow$: open switch, $\downarrow$: close switch.}
			\end{threeparttable}
			\label{Sol_switching_table_case4}%
		\end{table}%
	
		\begin{figure}[h!]
			\setlength{\abovecaptionskip}{0pt} 
			\setlength{\belowcaptionskip}{0pt} 
			\centering
			\includegraphics[width=0.49\textwidth]{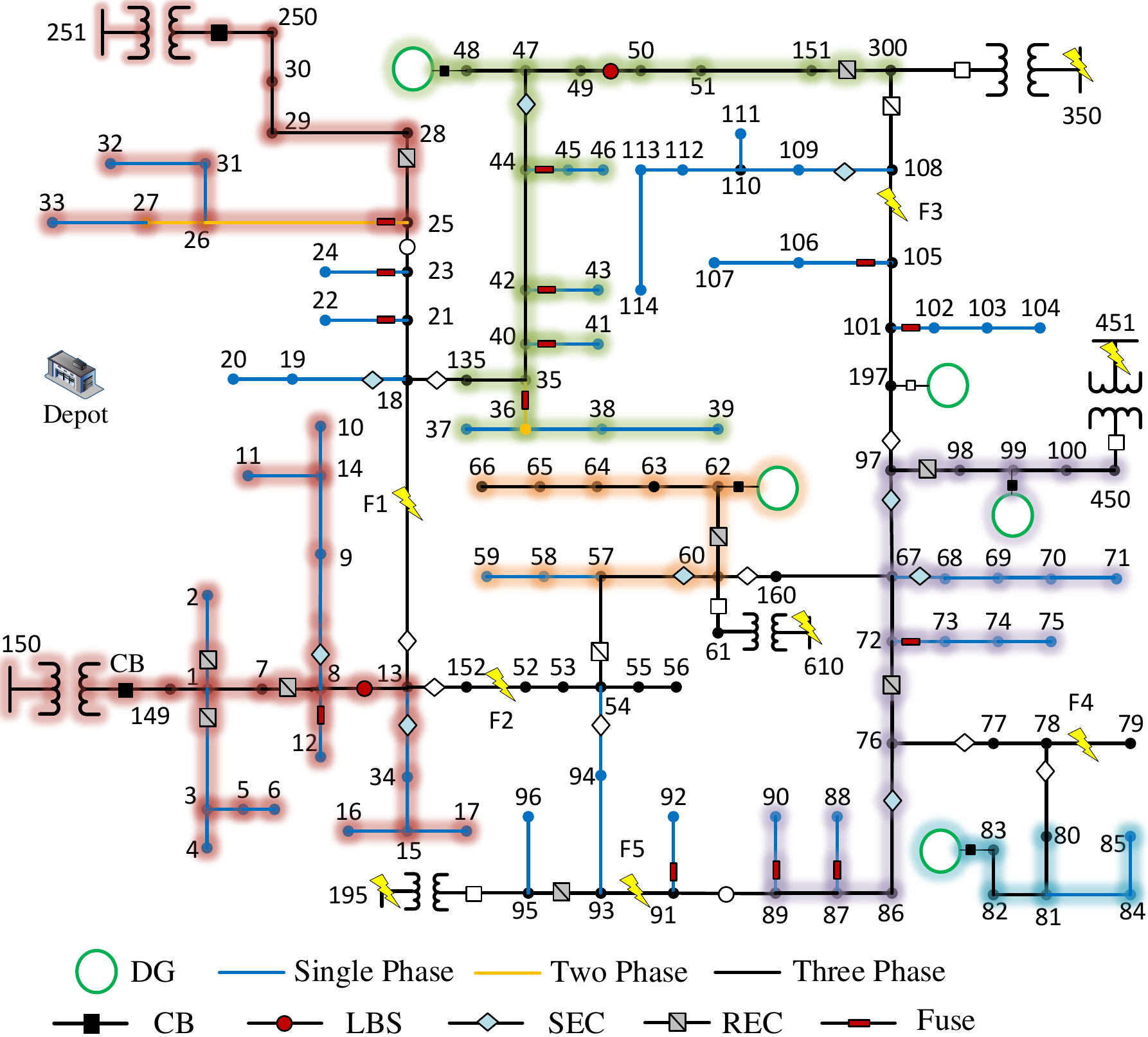}
			\caption{Final state of the network after sequential switching operations in test case IV.}\label{MG_test_2}
		\end{figure}
		
		\subsection{Test Case V}
		
		The final test case is conducted on the IEEE 8500-bus distribution system. The purpose of this case is to test the scalability of SSOP and its sensitivity to the number of steps. We modified the IEEE 8500-bus distribution system by adding switches and 4 DGs. A test case is simulated with 8 randomly selected damaged lines, as shown in Fig. \ref{8500IEEE}. The simulation is conducted with varying number of steps, starting from 0 to 40 steps. The result of the simulation is shown in Fig. \ref{Test_8500_step}, where the selected value for $|\Gamma|$ is 24 (using \eqref{|Gamma|}) and the computation time with $|\Gamma| = 24$ is 60 s. Therefore, the proposed method can be employed for large systems effectively. However, it is critical to select a proper number of steps. The problem is infeasible for $|\Gamma|$ less than 12 in this test case, and the computation time increases considerably with large numbers of steps, as shown in Fig. \ref{Test_8500_step}.
		
		\begin{figure}[h!]
			\setlength{\abovecaptionskip}{0pt} 
			\setlength{\belowcaptionskip}{0pt} 
			\centering
			\includegraphics[width=0.49\textwidth]{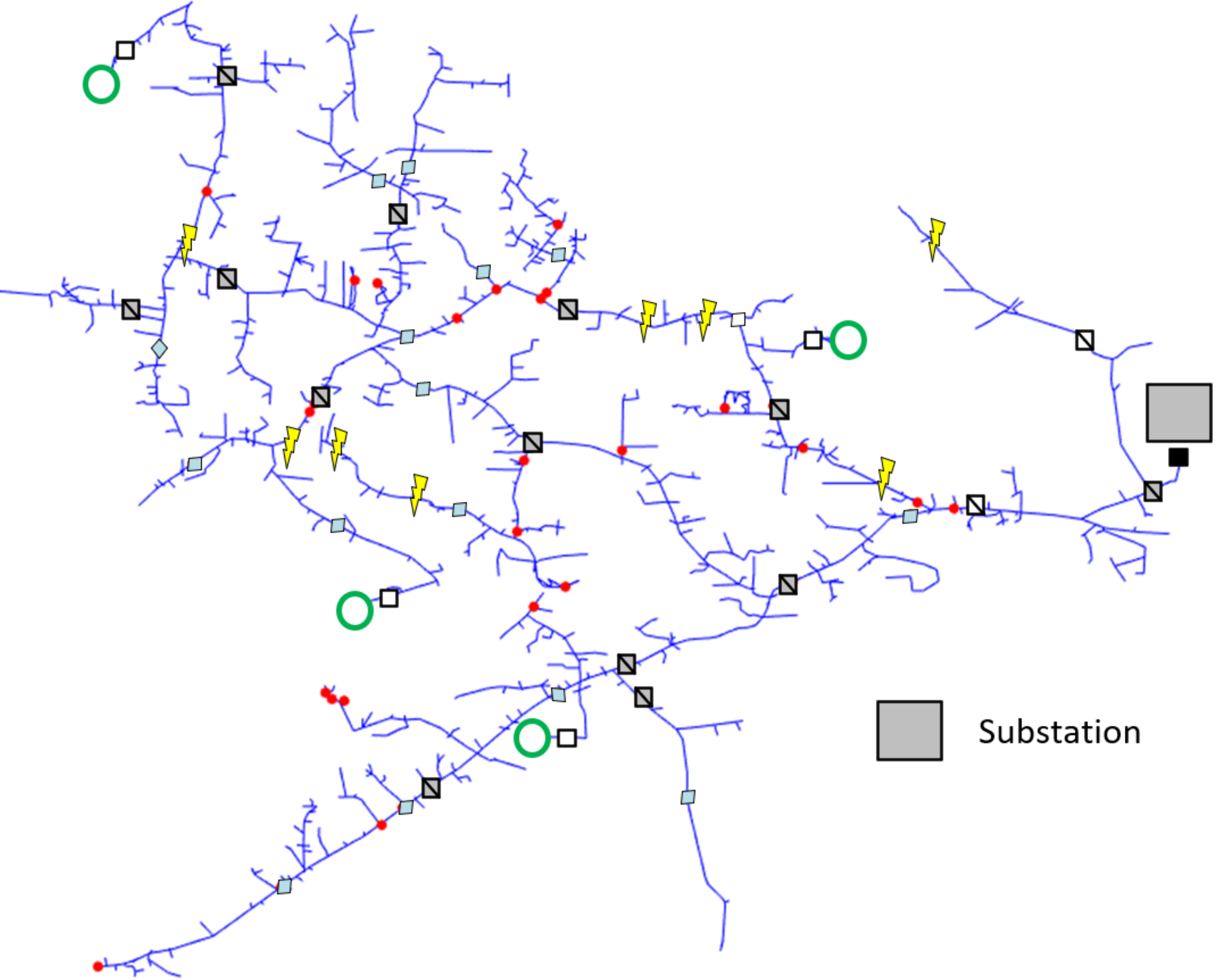}
			\caption{Initial state of the modified IEEE 8500-bus network with 8 damaged lines.}\label{8500IEEE}
		\end{figure}
		
		\begin{figure}[h!]
			\setlength{\abovecaptionskip}{0pt} 
			\setlength{\belowcaptionskip}{0pt} 
			\centering
			\includegraphics[width=0.49\textwidth]{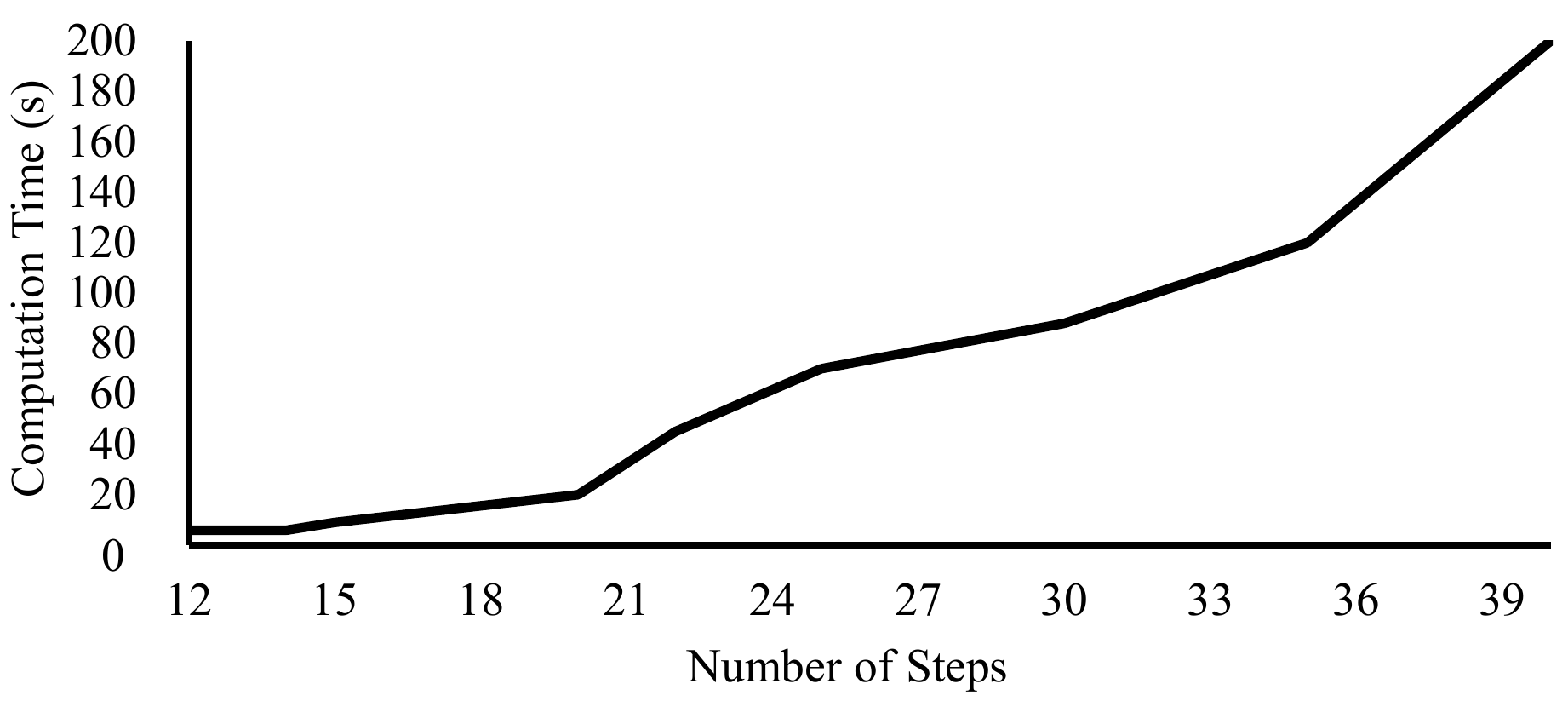}
			\caption{Sensitivity of the SSOP computation time with the change in number of steps for the IEEE 8500-bus system.}\label{Test_8500_step}
		\end{figure}
		
		\subsection{Discussion}
		
		As seen in the presented test cases, after faults are isolated, some of the unfaulted areas in the distribution network will experience an outage. The goal of the restoration problem is to reconfigure the network in order to supply these areas. The diversity of the switches, however, imposes a major challenge to this problem as the switches must be coordinated based on their characteristics. The results show that SSOP can perform sequential switching operations effectively, while adhering to the characteristics of the switches. Moreover, the model obtains the sequence of operation in an efficient time. Previous research assumed a uniform type of switch without limitations, which leads to infeasible solutions as shown in Table \ref{Sol_switching_table_case2_new} and Table \ref{Sol_switching_table_case4}. The test case on the IEEE 8500-bus system confirmed the scalability of the presented method, in addition to the importance of selecting a proper number of steps. The models presented in this paper can be important tools to assist distribution system operators in power restoration.
	}
	
	\section{Conclusion}\label{sec:8}
	
	We proposed an optimization strategy for distribution repair and restoration, while considering the characteristics of switching devices. {\color{black}Switches with constrained operational capabilities, such as SECs and LBSs, require special considerations when modeling network reconfiguration problems. 
	Once repair crews clear some of the faults, switches are operated to restore the cleared area while also isolating the remaining faults. Simulation results showed that the proposed method can effectively and efficiently find the required sequence of switching operations. The resulting switching operations highlight the importance of including the characteristics of the switches, as without them the switching sequence would be inapplicable in practice. The proposed SSOP model can be incorporated in future distribution network studies such as resilience and reliability planning.}
	
	\appendices
	{\color{black}
	\section{Optimal Topology Model}
	
	The mixed-integer linear programming formulation for the optimal topology problem is detailed below.
	\setcounter{equation}{0}
	\numberwithin{equation}{section}
	
	{\small
		\begin{equation}
		\textrm{min}~  \sum \limits_{\forall i \in \Omega_B}\big{(}(1-y_{i})  \rho^D_i \sum_{\forall \varphi} P^D_{i\varphi} \big{)}+ \sum_{\mathclap{\forall k \in \Omega_{SW}}} \rho^{SW}_k \gamma_{k}
		\label{Top_Obj}
		\end{equation}
		\begin{equation}
		(P_{k\varphi})^2 + (Q_{k\varphi})^2 \le (u_{k,}p_{k\varphi})(S_k)^2, \forall k \in \Omega_{K}, \varphi
		\label{Line_Limit_S} 
		\end{equation}
		\begin{equation}
		0 \le P_{i\varphi}^{G} \le \bar{P}_i^{G}, \forall i \in \Omega_{B},\varphi
		\label{DG limits_2}
		\end{equation}
		\begin{equation}
		0 \le Q_{i\varphi}^{G} \le \bar{Q}_i^{G},\forall i \in \Omega_{B},\varphi
		\label{QDG limits_2}
		\end{equation}
		\begin{equation}
		\mathop \sum_{\mathclap{\forall k \in \Omega_{K\left( {.,i} \right)}}} {P_{k\varphi}} + P^G_{i\varphi} + =\mathop \sum_{\mathclap{\forall k \in \Omega_{K\left( {i,.} \right)}}} {P_{k\varphi}} + P_{i\varphi}^D,\forall i \in \Omega_{B},\varphi
		\label{power balance_2}
		\end{equation}
		\begin{equation}
		\mathop \sum_{\mathclap{\forall k \in \Omega_{K\left( {.,i} \right)}}} {Q_{k\varphi}} + Q^G_{i\varphi} + =\mathop \sum_{\mathclap{\forall k \in \Omega_{K\left( {i,.} \right)}}} {Q_{k\varphi}} + Q_{i\varphi}^D,\forall i \in \Omega_{B},\varphi
		\label{reactive balance_2}
		\end{equation}
		\begin{equation}
		{\bm{U}_{j}} - {\bm{U}_{i}} + \bm{\bar{Z}}_k \bm{S}_{k}^*+  \bm{\bar{Z}}_k^*  \bm{S}_{k}\le (2-u_{k}-\bm{p}_{k}) M, \forall k \in \Omega_{K}
		\label{Top voltage1}
		\end{equation}
		\begin{equation}
		{\bm{U}_{j}} - {\bm{U}_{i}} + \bm{\bar{Z}}_k \bm{S}_{k}^*+  \bm{\bar{Z}}_k^*  \bm{S}_{k}\geq  -(2-u_{k}-\bm{p}_{k}) M, \forall k \in \Omega_{K}
		\label{Top voltage2}
		\end{equation}
		\begin{equation}
		\mathcal{X}_{i} \underline{U}  \le {{U}_{i\varphi}} \le \mathcal{X}_{i} \bar{U}, \forall i \in \Omega_{B}, \varphi
		\label{Top V limits}
		\end{equation}
		\begin{equation}
		2u_{k} \ge \mathcal{X}_{i} + \mathcal{X}_{j}, \forall k(i,j) \in \Omega_{F}
		\label{Top xx=0}
		\end{equation}
		\begin{equation}
		\mathcal{X}_{i} \ge y_{i}, \forall i \in \Omega_B
		\label{Top xx>y}
		\end{equation}
		\begin{equation}
		u_{k} = 1,  \forall k \in \Omega_K\backslash \{\Omega_{SW} \cup \Omega_{F}\}
		\label{Top uk=1}
		\end{equation}
		\begin{equation}
		u_{k} = \Gamma^0_k,  \forall k \in \Omega_{FS}
		\label{fuse_initial}
		\end{equation}
		\begin{equation}
		\gamma_{k} \ge u_{k}-\Gamma^0_k, \forall k \in \Omega_{SW}
		\label{Top sw1}
		\end{equation}
		\begin{equation}
		\gamma_{k} \ge \Gamma^0_k-u_{k}, \forall k \in \Omega_{SW}
		\label{Top sw2}
		\end{equation}
	}
		
		The first term in objective (\ref{Top_Obj}) minimizes the cost of load shedding, while the second term minimizes the cost of operating the switches. The limits on the line-flow constraints in (\ref{Line_Limit_S}) is multiplied by $u_{k}$ so that if a line is damaged or a switch is opened, there will be no power flowing on it. If line $k(i,j)$ connecting buses $i$ and $j$ is two-phase (e.g., phases $a$ and $b$), then power can only flow on these two phases, which is realized by including ${p_{k\varphi}}$. Constraint (\ref{Line_Limit_S}) is linearized using the circular constraint linearization method presented in \cite{Zhao2019}. Constraints (\ref{DG limits_2}) and (\ref{QDG limits_2}) represent the active and reactive power limits for the generators/substations, respectively. The power balance constraints are formulated in (\ref{power balance_2}) and (\ref{reactive balance_2}). We adapt the formulation in \cite{Gan2014} to model the unbalanced power flow equations. Constraints (\ref{Top voltage1})--(\ref{Top voltage2}) represent Kirchhoff's voltage law (KVL), where $\bm{U}_i$ is a vector representing the three-phase voltages ($[|V_i^a|^2,|V_i^b|^2,|V_i^c|^2]^T$), and $\bm{\bar{Z}}_{k}$ is the impedance of line $k$ multiplied by a phase shift matrix \cite{Gan2014}. The big $M$ method is used to decouple the voltages between lines that are disconnected or damaged in (\ref{Top voltage1}) and (\ref{Top voltage2}). Constraint (\ref{Top V limits}) ensures that the voltage is within a specified limit, and 0 if the bus is in an outage area. Constraint (\ref{Top xx=0}) sets the values of $\mathcal{X}_i$ and $\mathcal{X}_j$ to 0 if line $k$ is damaged. Constraint (\ref{Top xx>y}) states that if bus $i$ is de-energized, then the load must be shed. Constraint (\ref{Top uk=1}) defines the default status of the lines that are not damaged or not switchable and constraint (\ref{fuse_initial}) sets the status of the fuses. Constraint (\ref{Top sw1})--(\ref{Top sw2}) determine the switching operation status ($\gamma_{k}$). In addition to the above constraints, we impose radiality using the formulation in \cite{Arif2018c}.  
		
		{\color{black}
			
			\section{Repair Crew Routing Model}
		}
		In RCRP, crews are dispatched to the distribution system in order to repair the damaged components. A crew's path is determined by the variable $\grave{x}_{ijc}, i \in \hat{\Omega}_{DB}, j \in \hat{\Omega}_{DB}$, where $\grave{x}_{ijc} = 1$ if crew $c$ travels from bus block $i$ to $j$. Once a crew reaches a bus block, it is assigned to the damaged components inside the bus blocks using $\mathcal{W}_{kc}, k \in \Omega_{F(i)}$, where $\Omega_{F(i)}$ is the set of damaged lines in bus block $i$. The RCRP model is formulated below:
		
		{\small
			\begin{equation}
			\textrm{min}~  \sum \limits_{\forall i \in \Omega_{BL}}\big{(}(1-y_{i})  \rho^D_i \sum_{\forall \varphi} \tilde{P}^D_{i\varphi} + \rho^T_{ji}\sum \limits_{\forall j\in \hat{\Omega}_{BL}}\sum \limits_{\forall c}tr_{ji}\grave{x}_{jic}\big{)}
			\label{Crew_Obj}
			\end{equation}
			\begin{equation}
			\sum_{\forall i \in \hat{\Omega}_{DB}}\sum_{\forall c} \grave{x}_{ijc} \ge 1, \forall j \in {\Omega}_{DB}
			\label{visit}
			\end{equation}
			\begin{equation}
			\mathop \sum \limits_{\forall i \in \hat{\Omega}_{DB}}{\grave{x}_{0ic}} = 1, \forall c
			\label{start_from_depot}
			\end{equation}
			\begin{equation}
			\mathop \sum \limits_{\forall i \in \hat{\Omega}_{DB}} {\grave{x}_{i 0 c}} = 1, \forall c
			\label{back_to_depot}
			\end{equation}
			\begin{equation}
			\mathop \sum_{{\forall j \in \hat{\Omega}_{DB}\backslash \left\{ i \right\}}}
			{\grave{x}_{ijc}} - \mathop \sum_{{{\forall j \in \hat{\Omega}_{DB}\backslash \left\{ i \right\}}}} {\grave{x}_{jic}} = 0,\forall c,i \in {\Omega}_{DB}
			\label{path_flow}
			\end{equation}
			\begin{equation}
			\sum_{\forall c} \mathcal{W}_{kc} = 1, \forall i \in {\Omega}_{DB}, k \in {\Omega}_{F(i)}
			\label{one_job}
			\end{equation}
			\begin{equation}
			\sum_{\forall k \in {\Omega}_{F(i)}} \mathcal{W}_{kc} \le |{\Omega}_{F(i)}| \sum_{\forall j \in \hat{\Omega}_{DB}} \grave{x}_{ijc}, \forall i \in {\Omega}_{DB}, c
			\label{must_visit_area}
			\end{equation}
			\begin{equation}
			\begin{split}
			\grave{\alpha}_{ic} + \sum_{\forall k \in {\Omega}_{F(i)}}& {ET_{kc}} \mathcal{W}_{kc} + \grave{tr_{ij}} - \left( {1 - \grave{x}_{ijc}} \right)M \\ \le \grave{\alpha}_{jc},~
			& \forall i \in \hat{\Omega}_{DB},j \in \Omega_{DB}, i \neq j, c
			\end{split}
			\label{Arrival_const}
			\end{equation}
			\begin{equation}
			\mathcal{R}_{i} \ge \grave{\alpha}_{ic} + \sum_{\forall k \in {\Omega}_{F(i)}} {ET_{kc}} \mathcal{W}_{kc}, \forall i \in {\Omega}_{DB}, c
			\label{Rest_const}
			\end{equation}
			\begin{equation}
			t~(1-x^F_{it}) + M x^F_{it} \ge \mathcal{R}_{i}, \forall i \in {\Omega}_{DB}, t
			\label{crew_fault_state}
			\end{equation}
			\begin{equation}
			u_{kt} = (1-x^F_{it}), \forall k \in \Omega_{MF},  i \in \Omega_{DB(k)}, t
			\label{fuse_rep}
			\end{equation}
			\begin{equation}
			u_{k 0} = \Gamma^F_k, \forall k \in \Omega_{SW}
			\label{crew_initial_u}
			\end{equation}
			\begin{equation}
			-(1-u_{kt}) \le x^F_{it} - x^F_{jt} \le (1-u_{kt}), \forall k(i,j) \in \Omega_{SW}, t
			\label{Sec_propagate_faultB}
			\end{equation}
			\begin{equation}
			y_{it} \le 1-x^F_{it}, \forall i \in \Omega_{BL}, t
			\label{Sec_y_xFB}
			\end{equation}
			\begin{equation}
			0 \le P_{i\varphi t}^G \le \bar{P}^G, \forall i \in \Omega_{BL}, \varphi, t
			\label{SSOP_PG_limitB}
			\end{equation}
			\begin{equation}
			-\bar{Q}^G \le Q_{i\varphi t}^G \le \bar{Q}^G, \forall i \in \Omega_{BL}, \varphi, t
			\label{SSOP_QG_limitB}
			\end{equation}
			\begin{equation}
			P_{i\varphi t}^G + \sum_{\mathclap{\forall k \in K(.,i)}}P_{k\varphi t} = y_{it} \tilde{P}_{i\varphi}^D+ \sum_{\mathclap{\forall k \in K(i,.)}}P_{k\varphi t}, \forall i \in \Omega_{BL}, \varphi, t
			\label{SSOP_P_flowB}
			\end{equation}
			\begin{equation}
			Q_{i\varphi t}^G + \sum_{\mathclap{\forall k \in K(.,i)}}Q_{k\varphi t} = y_{it} \tilde{Q}_{i\varphi}^D+ \sum_{\mathclap{\forall k \in K(i,.)}}Q_{k\varphi t}, \forall i \in \Omega_{BL}, \varphi, t
			\label{SSOP_Q_flowB}
			\end{equation}
			\begin{equation}
			P_{k\varphi t}^2+Q_{k\varphi t}^2 \le u_{k t} p_{k \varphi} \bar{S}_k^2, \forall k \in \Omega_{SW}, \varphi, t
			\label{SSOP_S_limitB}
			\end{equation}
		}
	\noindent
		The first and second terms in (\ref{Crew_Obj}) minimize load shedding and the distance traveled by the crews, respectively. Constraint (\ref{visit}) indicates that each damaged bus block must be visited by at least one crew. 
		Constraints (\ref{start_from_depot})--(\ref{back_to_depot}) define the starting and ending locations for the repair crews. Equation (\ref{path_flow}) represents the path-flow constraint for the routing problem. 
		Each damaged component is assigned to one crew in constraint (\ref{one_job}). For $k \in \Omega_{F(i)}$, crew $c$ is assigned to damaged component $k$ only if the crew visits bus block $i$, this is enforced by (\ref{must_visit_area}). Constraint (\ref{Arrival_const}) defines the arrival time of each crew at the damaged bus blocks, such that $\grave{\alpha}_{jc}$ equals the sum of $\grave{\alpha}_{ic}$, travel time between $i$ and $j$, and the time spent at the bus block. Constraint (\ref{Rest_const}) defines the time when the bus block is repaired. A bus block is repaired once all damaged components in the area are repaired. The value of $x^F_{it}$ (damage state) is determined in (\ref{crew_fault_state}), where $x^F_{it}=0$ for $t \ge \mathcal{R}_i$. For a bus block that is connected to a melted fuse, the last crew to leave the bus block will replace the fuse. The statuses of fuses are determined by (\ref{fuse_rep}), where $i \in \Omega_{DB(k)}$ is the bus block protected by fuse $k$. Constraint (\ref{crew_initial_u}) defines the initial state of the switches, where the initial state of RCRP is the final state of OTP ($\Gamma^F_k$). Constraint (\ref{Sec_propagate_faultB}) models the propagation of faults between connected bus blocks. Constraint (\ref{Sec_y_xFB}) states that a faulted bus block cannot be served. The combination of (\ref{Sec_propagate_faultB}) and (\ref{Sec_y_xFB}) ensures that the faults must be isolated to serve the loads. The active and reactive power generation limits are given in (\ref{SSOP_PG_limitB}) and (\ref{SSOP_QG_limitB}), respectively. The power balance constraints are given in (\ref{SSOP_P_flowB}) and (\ref{SSOP_Q_flowB}). Constraint (\ref{SSOP_S_limitB}) models the line thermal limit. In addition, radiality is enforced using the spanning tree constraints in \cite{Arif2018c}.

}

		\vspace{-0.8cm}
				\begin{IEEEbiography}[{\includegraphics[width=1in,height=1.25in,clip]{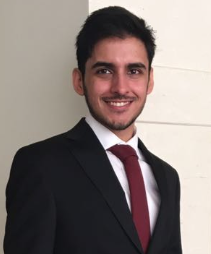}}]{Anmar Arif}
				(M'19) is an Assistant Professor at King Saud University and a Research Associate at the University of Manchester. He received his Ph.D. in Electrical and Computer Engineering from Iowa State University, Ames, IA, USA. He received his B.S. and MSE degrees in electrical engineering from King Saud University and Arizona State University in 2012 and 2015, respectively. He was Academic Visitor at the University of Oxford from 2019 to 2021. His research interest includes power system optimization, outage management, operations research, transportation systems, and machine learning.
				\end{IEEEbiography}
			
			\vspace{-0.8cm}
			
				\begin{IEEEbiography}[{\includegraphics[width=1in,height=1.25in,clip]{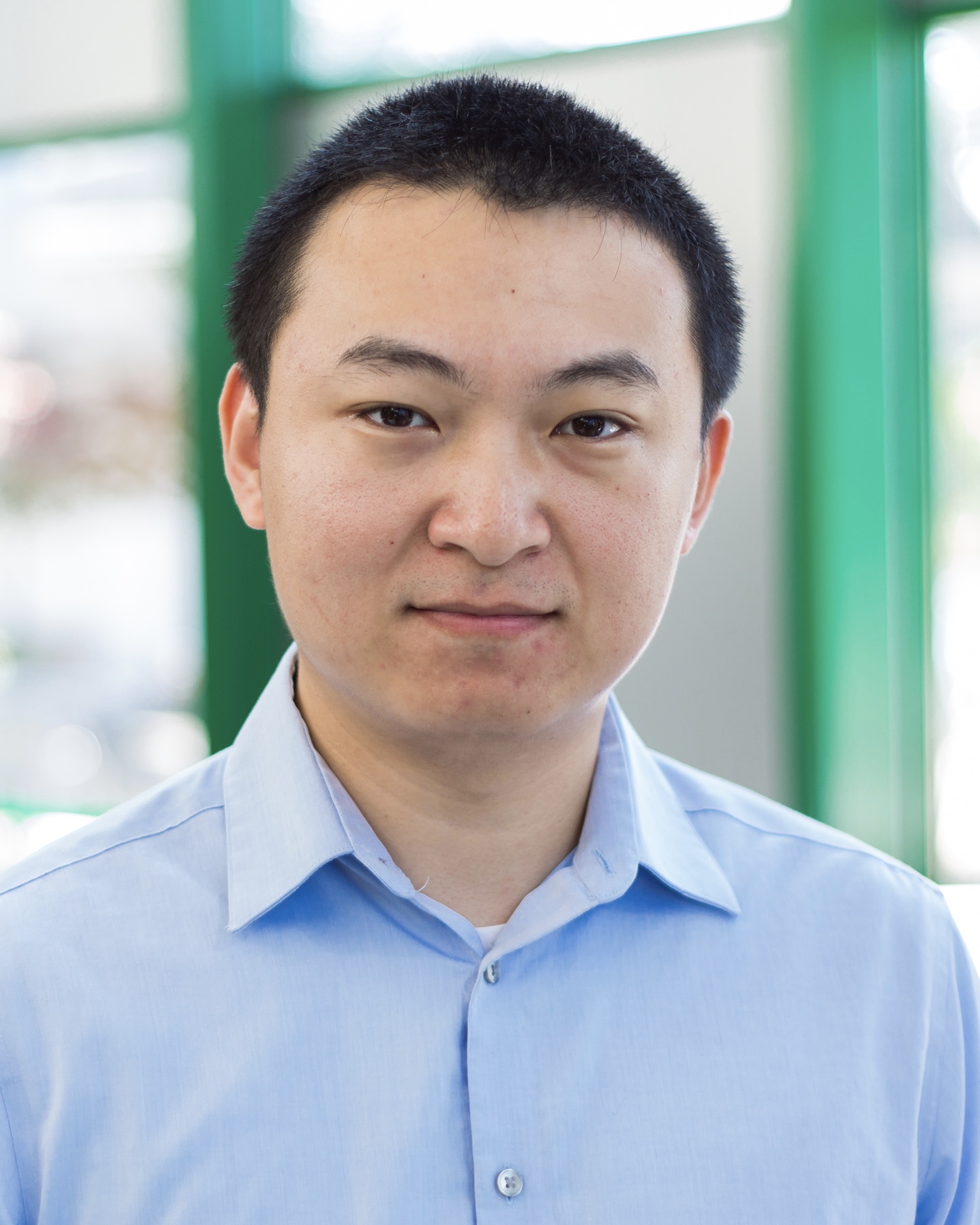}}]{Bai Cui}
				(M'18) received the first B.S. degree in electrical engineering from Shanghai Jiao Tong University, Shanghai, China, the second B.S. degree in computer engineering from the University of Michigan, Ann Arbor, MI, USA, in 2011, and the Ph.D. degree in electrical engineering from the Georgia Institute of Technology, Atlanta, GA, USA, in 2018. He was a Post-Doctoral Appointee with Argonne National Laboratory, Lemont, IL, USA, from 2018 to 2019. He is currently a Post-Doctoral Researcher with the National Renewable Energy Laboratory, Golden, CO, USA. His research focuses on the control and optimization of power systems, with an emphasis on security assessment and renewable integration.
				\end{IEEEbiography}
				
				\vspace{-0.7cm}
				
				\begin{IEEEbiography}[{\includegraphics[width=1in,height=1.25in,clip]{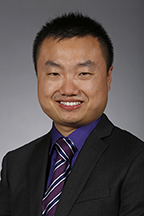}}]{Zhaoyu Wang}
				(SM'20) received the B.S. and M.S. degrees in electrical engineering from Shanghai Jiaotong University, and the M.S. and Ph.D. degrees in electrical and computer engineering from Georgia Institute of Technology. He is the Harpole-Pentair Assistant Professor with Iowa State University. His research interests include optimization and data analytics in power distribution systems and microgrids. He was the recipient of the National Science Foundation CAREER Award, the IEEE Power and Energy Society (PES) Outstanding Young Engineer Award, College of Engineering?s Early Achievement in Research Award, and the Harpole-Pentair Young Faculty Award Endowment. He is the Principal Investigator for a multitude of projects funded by the National Science Foundation, the Department of Energy, National Laboratories, PSERC, and Iowa Economic Development Authority. He is the Chair of IEEE PES PSOPE Award Subcommittee, the Co-Vice Chair of PES Distribution System Operation and Planning Subcommittee, and the Vice Chair of PES Task Force on Advances in Natural Disaster Mitigation Methods. He is an Associate Editor of IEEE TRANSACTIONS ON POWER SYSTEMS, IEEE TRANSACTIONS ON SMART GRID, IEEE OPEN ACCESS JOURNAL OF POWER AND ENERGY, IEEE POWER ENGINEERING LETTERS, and IET Smart Grid.
				\end{IEEEbiography}

\end{document}